%% file: arxiv_0628llm4design.tex
\UseRawInputEncoding
\documentclass[sigconf]{acmart}

\AtBeginDocument{%
  }
    
\setcopyright{acmlicensed}
\copyrightyear{2018}
\acmYear{2018}
\acmDOI{XXXXXXX.XXXXXXX}

\usepackage{xcolor} 
\usepackage{listings} 

\usepackage{graphicx} 
\usepackage{booktabs}
\usepackage{tabularx} 
\usepackage{lipsum} 
\usepackage{graphicx}
\begin{document}


\title{LLM4DESIGN: An Automated Multi-Modal System for Architectural and Environmental Design }


\author{Ran Chen}
\authornote{Both authors contributed equally to this research.}
\email{chenran705367787@bjfu.edu.cn}
\orcid{0000-0001-9875-4588}
\author{Xueqi Yao}
\authornotemark[1]
\email{Yxq18210187099@bjfu.edu.cn}
\affiliation{%
  \institution{Beijing Forestry University}
  \streetaddress{Haidian}
  \city{Beijing}
  \country{China}}

\author{Xuhui Jiang}
\email{jiangxuhui19g@ict.ac.cn}
\authornote{Corresponding Author}
\affiliation{%
 \institution{Institute of Computing Technology, Chinese Academy of Sciences}
 \streetaddress{Haidian}
 \city{Beijing}
 \country{China}}





\renewcommand{\shortauthors}{Trovato et al.}

\begin{teaserfigure}
  \includegraphics[width=\textwidth]{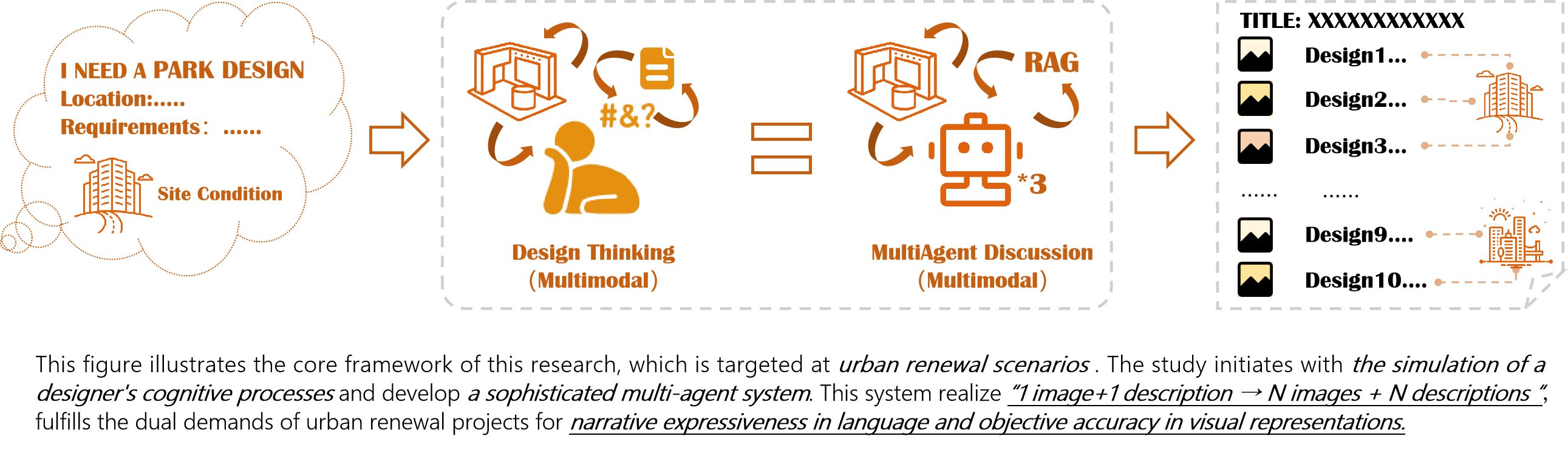}
  \Description{Our AI designer}
  \label{fig:teaser}
\end{teaserfigure}

\begin{abstract}

This study introduces LLM4DESIGN, a highly automated system for generating architectural and environmental design proposals. 
LLM4DESIGN, relying solely on site conditions and design requirements, employs Multi-Agent systems to foster creativity, Retrieval Augmented Generation (RAG) to ground designs in realism, and Visual Language Models (VLM) to synchronize all information. This system resulting in coherent, multi-illustrated, and multi-textual design schemes. 
The system meets the dual needs of narrative storytelling and objective drawing presentation in generating architectural and environmental design proposals. 
Extensive comparative and ablation experiments confirm the innovativeness of LLM4DESIGN's narrative and the grounded applicability of its plans, demonstrating its superior performance in the field of urban renewal design.
Lastly, we have created the first cross-modal design scheme dataset covering architecture, landscape, interior, and urban design, providing rich resources for future research.

\end{abstract}



\received{20 February 2007}
\received[revised]{12 March 2009}
\received[accepted]{5 June 2009}

\maketitle

\section{Introduction}

Architectural and environmental design must balance narrative language with objective descriptions. Engaging stories can make design proposals more attractive and easier to market, but without precise objective descriptions, these stories cannot be realized since designs must be constructed in the real world. Therefore, a careful balance of both is crucial to ensure the design is both imaginative and feasible.In the process of AI assistance in architectural and environmental design, there are three main issues that need to be addressed. These issues directly reflect the challenges of balancing narrative language with objective descriptions.

\begin{figure*}
  \centering  
  \includegraphics[width=\textwidth]{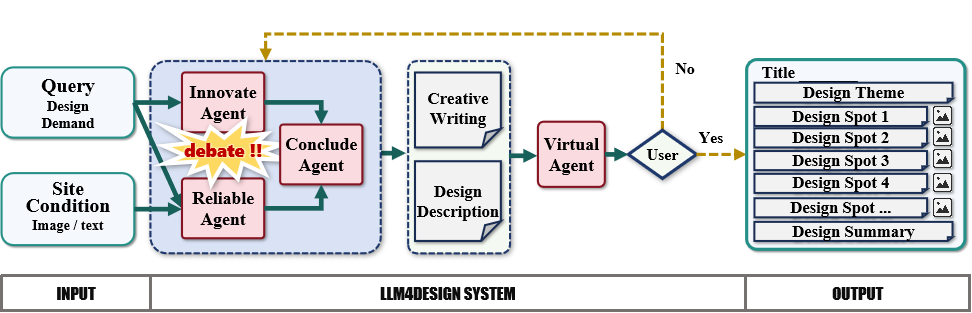}  
  \caption{Main Workflow}  
  \Description{A description of the image}  
\end{figure*}

\textbf{Challenge1: Design documents must include compelling creative stories.} Architects and environmental designers need to create stories that are not only engaging but also rooted in the practical realities of the site during case studies, site analysis, and scheme development. The first challenge in using AI for architectural and environmental design is developing AI that can narrate these stories from a designer's perspective while ensuring they align with real-world feasibility.

\textbf{Challenge2: Design documents must reflect real-life situations.} Design documentation must account for the actual limitations and potential of the site, ensuring innovative and applicability. The second challenge for AI in architectural and environmental design is how to maintain innovation while ensuring that the plans are suitable for the site's actual conditions.

\textbf{Challenge3: Design documents require cross-modal information alignment.} All design strategies need to be accurately reflected in site images to ensure the practical executability of the design. This requires attractive stories and visual descriptions both contained in the design scheme, reinforcing the feasibility and implementability of the design.

\textbf{To address these challenges,} we have designed a comprehensive framework aimed at stimulating\textbf{ "innovative" thinking i}n large models while ensuring the \textbf{"reliability" }of these thoughts. The framework extends through an image generation model based on textual modality to align information across various modalities. Ultimately, this will generate a multimodal design proposal that users can further adjust.

Guided by common design workflows and grounded in theoretical methods of design thinking, our system comprises the following components: Multimodal Retrieval; Multiagent Simulation; Visualization Expression; User Evaluation.

\textbf{To be more specific:} \textbf{In response to Challenge 1,} we investigated existing research related to creativity in large models and found that beneficial illusions present in these models can serve as sources of creativity \cite{jiang2024survey, franceschelli2023creativity}. Methods such as prompt engineering and agents have been proven to effectively stimulate creativity. Moreover, using multiple agents can circumvent the limitations of iterative thought processes \cite{sel2023algorithm}. Therefore, to maximize the creative potential of large models, we have chosen to utilize multiagent systems. This paper focuses on the construction of multiagents as the core mechanism to simulate the essential nature of design thinking. Meanwhile, We drew inspiration from the adversarial generation, setting different roles to complete creative writing.

\textbf{In response to Challenge 2,}  we have integrated the Remote Associates Test (RAT) into the discussion process of the AI agents. During the imaginative phase of design, the agents continuously perform searches, using cases to support and provide feedback on their thought processes.

\textbf{In response to Challenge 3,} we have designed the Conclude Agent. This agent is constructed based on the Visual Language Large Model (VLLM). It organizes the previously created creative narratives, retaining the story while distilling design descriptions suitable for graphical representation.

\textbf{As for the evaluation method }, design is an open and complex issue, and we cannot define a "good" design with a single metric.
Our goal is to be as \textbf{"Innovative"} as possible within the \textbf{"Reliable"} range, aiming to find the limit of how far from \textbf{"Reliable"} we can go.
\textbf{Thus, we also draw from adversarial generation concepts to design the evaluation metric "balance," allowing two Agent frameworks to grow through competition and achieve Nash equilibrium.}
However, if balance is the sole metric, the algorithm's goal would converge to an optimal solution, which is undesirable in design, making balance only a reference metric.
Design does not seek a unique solution, and different designers have different preferences. Thus, we use \textbf{"Innovative"} and \textbf{"Reliable"} as two reference metrics to form a comprehensive evaluation system for different algorithms' performance.
Finally, recognizing that text alone can't fully convey design information for effective use in architectural design, we integrated an essential image control generator, the virtual agent, into our framework to address cross-modal design challenges comprehensively.

In summary, the contributions of this work are as follows:

\textbf{Contribution 1: Multi-agent system inspires creativity.} The multi-agent system uses different designers' perspectives to stimulate creativity. This setup allows agents to work concurrently, exploring diverse design solutions and enhancing innovation and design variety.

\textbf{Contribution 2: Combining RAG to ensure accuracy.}The AI system seamlessly integrates Retrieval-Augmented Generation, leveraging existing design data to enrich and enhance creative content. This approach guarantees that the generated designs are innovative and highly accurate, satisfying the demands and constraints of various practical scenarios.

\textbf{Contribution 3: Using visual language models to align strategy information with site data.} Visual language models allow the AI to comprehend and interpret site visuals, aligning them with design strategies effectively. This approach ensures that design plans are both accurate and aligned with the site's specific characteristics, improving design coherence and feasibility.

\section{Related Work}

AI technologies have deeply penetrated the design industry. According to \cite{aiadsurvey}, AI-assisted design primarily encompasses four directions: discovery, visualization, creation, and testing. In the "discovery" phase, AI tools \cite{aiad_discover1, aiad_discover2, aiad_discover3} analyze user data to reveal underlying needs and guide preliminary design ideas. During the "visualization" phase, AI aids designers in precisely constructing design concepts. In the "creation" phase, AI automatically drafts and iteratively refines design schemes. In the "testing" phase, AI evaluates the design quality to ensure the practicality and customer satisfaction of the outcomes.

Therefore, the current AI-assisted design (AIAD) process is still performed in stages. However, the complexity and non-linear characteristics of the design process demand a more advanced integration strategy to balance innovation with practicality and meet unique site and client needs.

This section reviews the applications of AI in the fields of image and text generation. It finds that while image generation focuses on control and often lacks preliminary design reasoning, text generation is rich in creativity but lacks specificity. Consequently, this study develops a multimodal framework by integrating the strengths of both image and text generation. This framework aims to mimic the thought process of designers and achieves a one-click operation from "user query" to "multimodal drawing generation," enhancing the automation and precision of design.

\subsection{Image Generation}
Recent advancements in image generation technology, particularly through Generative Adversarial Networks (GANs) \cite{nauata2020house,quan2022urban}, have significantly enhanced the methods available to architects and landscape designers, enabling the generation of diverse building and landscape floor plans based on user inputs. However, these technologies have primarily been applied to floor plan generation\cite{wang2021actfloor}. The introduction of large models such as diffusion and VAEs has lowered technical barriers  and enabled preliminary cross-modal generation. Despite these advancements, text-to-image processes still struggle with inadequate control over outputs, resulting in imprecise details and requiring extensive prior knowledge.

Current research in this field primarily focuses on precise generation. However, in industrial workflows, control over image generation is often deferred to the final drawing stages, with earlier stages still lacking sufficient tools to systematically reason through design ideas and comprehensively evaluate their effectiveness. 

\subsection{Creative Text Generation}

In the current landscape of artificial intelligence research and applications, agents have demonstrated extensive potential and diverse functional capabilities\cite{agentsurvey}. 

These agents have been effectively applied in areas such as embodied intelligence \cite{jushen1,jushen2,jushen3}, tool use \cite{tool1,tool2,tool3,tool4}, and social simulation \cite{shehuimoni,shehuimoni1}, and have also penetrated vertical domains, including psychology \cite{xinlixue}, law \cite{chatlaw}, and chemistry \cite{chemistry}. 

Despite this, the field of design, especially tasks related to architectural and landscape design, has yet to see significant applications of agents. However, technologies such as prompt engineering and chain of thought reasoning \cite{prompt_hubert2023artificial,prompt_koivisto2023best} have proven to be exceptionally effective in stimulating innovative capabilities in these areas.

\section{Methodology}
In this section, we will first illustrate the meanings of \textbf{"design languages"} discussed previously, through a case study. This is to intuitively show why it is necessary to perform information retrieval and why it is crucial to filter and summarize beneficial design illusions. Subsequently, we will elaborate on each component of the framework in detail.

\subsection{"Design Languages"}

In the preliminary phase of our research, we conducted comparative experiments to assess the feasibility of directly employing existing large-scale models to generate answers based on user requirements.

The results indicated significant inaccuracies in the text generated directly, leading to responses that deviated from reality and struggled to produce reasonable solutions based on available information, failing to offer authentic design proposals.

Taking the Suzhou Garden Museum in China as an example(Figure 2), designers creatively reinterpreted "the spirit of China's traditional history" as "The black and white," "the museum's theme of 'books,'" as "The elegantly staggered facade," and "the theme of landscape paintings" as "sketching in black and white," whereas existing large models tend to translate these more directly, such as turning "the museum's theme of 'books,'" into "sculptures of opened books and scrolls.

We discovered that the core issue with models encountering problems in unconstrained conversational contexts lies in the significant difference in handling design-related semantic information as opposed to processing conventional natural language texts.

The complexity and specificity of design semantics require models to not only understand standard linguistic expressions but also possess in-depth understanding and processing capabilities for specialized knowledge in the design field.

For instance, the language used by designers is categorized under natural language. However, it frequently incorporates industry-specific terminology and concepts, such as "historical sentiment," which are difficult for traditional natural language processing models to accurately interpret.

Additionally, designers frequently refer to "scene memories" to describe design concepts, which in visualization often correspond to fluid curves or specific shapes, details that are also difficult to capture in traditional text processing.

\begin{figure}[H]
    \centering
    \includegraphics[width=\linewidth]{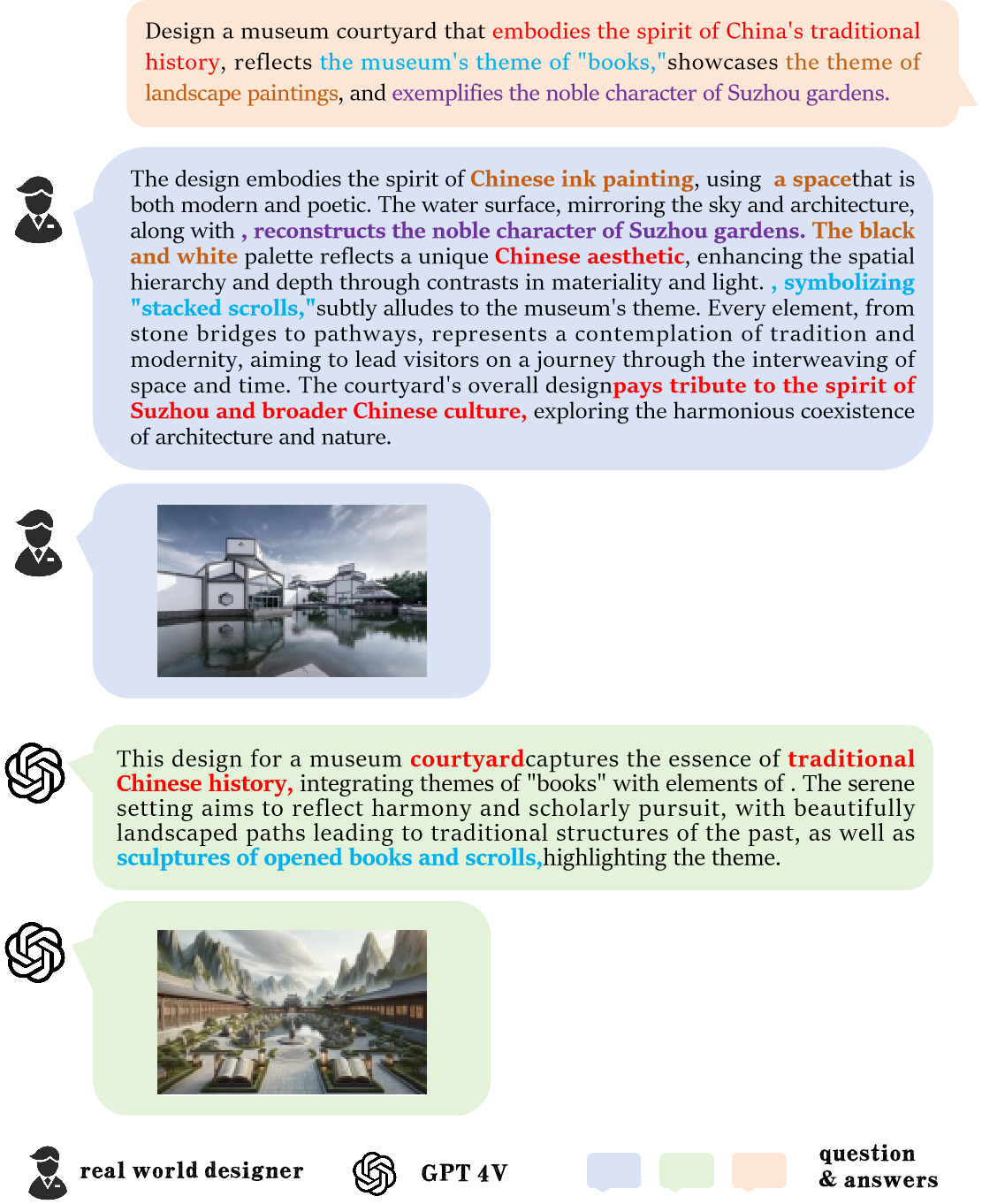}
    \caption{A case of design language}
    \Description{An image showing a design language example with various elements highlighted.}
    \label{1}
\end{figure}

\subsection{Design Language Aligner}

Based on the analysis in Section 3.2, we conclude that a primary challenge with the current agents' divergent capabilities is the gap between design language and natural language. 

Therefore, we developed a converter. It aims at aligning design language with natural language.The construction of the aligner consists of the following steps:

\textbf{Fine-tuning the BGE-M3 model}: The BGE-M3 model is primarily used for text-to-text search. In evaluating the performance of various text embedding models, we found that the Flag Embedding model excelled in multiple metrics. We fine-tuned the Flag Embedding model with 5k high-quality data entries from a multimodal database, with specific parameters detailed in the table below.

\textbf{Fine-tuning the BLIP2 model}: The BLIP2 model is mainly used for text-to-image search. We fine-tuned the BLIP2 model using a specific dataset to enhance its capability in searching and recognizing design images.

\textbf{Constructing a vector database}: We collected 60k data to establish a multimodal vector database.

\subsection{Debate between {Agent\_I} and {Agent\_R}}

In the preliminary stage of the research, it is imperative to clearly define two categories of agents: the Innovative Agent (Agent\_I) and the Retrieval Agent (Agent\_R).

\begin{itemize}

\item The \textbf{Agent\_I} represents agents with a high capacity for innovation. We employ methods to stimulate remote associations, encouraging them to imagine the design scheme in depth and breadth, thereby maximizing their creative potential.

\item The \textbf{Agent\_R}, on the other hand, is an agent based on retrieval, whose utterances must rely on a pre-constructed library of design cases. This library, comprising various design examples, serves as a crucial tool for evaluating the convergence of the design scheme.
\end{itemize}

\subsection{The conclusion of {Agent\_C} }

The Conclude Agent represents a pivotal component of our AI-driven design framework, developed to refine and synthesize design narratives. It is founded on the Visual Language Large Model (VLLM), a cutting-edge model adept at processing and generating multimodal content. The primary function of the Conclude Agent is to streamline and coalesce the creative narratives generated in earlier phases of the design process. This agent meticulously preserves the essence of the story while concurrently extracting and condensing the design elements necessary for visual depiction. By doing so, it ensures that the resulting graphical representations remain coherent with the original narrative, effectively bridging the gap between textual creativity and visual expression.

\subsection{Graphic Align and represent by {Agent\_V}}

Agent\_V, known as the Visual Agent, is primarily tasked with transforming design thinking into graphical representations. Its core responsibilities include:Receiving prompts from Agent\_C and conducting secondary searches to locate reference images based on these prompts;Matching prompts with reference images using "design keywords ; Generating images by integrating prompts with ControlNet \cite{controlnet} ; Compiling the outputs, presenting them to the user, and conducting quality reviews.

In the study of text-to-image generation, after comparing existing state-of-the-art models, we have chosen StableDiffusion as our foundational model for image generation. 

Additionally, ControlNet\cite{controlnet} has been proven to be an effective strategy for controlling site-specific information, primarily utilizing linear  and segmentation   methods.The parameters we used for graphical representation are as followed.

\begin{table}[H]
  \caption{Parameter of Stablediffusion}
  \label{tab:results}
  \centering
  \begin{tabularx}{\columnwidth}{>{\centering\arraybackslash}X >{\centering\arraybackslash}X >{\centering\arraybackslash}X >{\centering\arraybackslash}X}
    \toprule
    Model &  Checkpoint & ControlNet & Lora \\
    \midrule
    SDV1.5 &  Landscape SuperMix & Linear & / \\
    \bottomrule
  \end{tabularx}
\end{table}

\section{Experiment Settings}

\subsection{Datasets}
In the dataset construction phase, this study established the inaugural multimodal dataset tailored for the landscape architecture and planning field. Given the complexity and varied formats of design texts, web scraping techniques were employed to selectively extract and cleanse data from authoritative journals within the domain. We utilized GPT-3.5-turbo api to generate pairs of textual queries and design descriptions. Image captions were produced using the LLaVA algorithm. The final dataset is structured as follows, comprising 65,332 entries, which can facilitate a wide range of multimodal research within this field.

\begin{itemize}
    \item ID: The unique identifier for each data entry.
    \item IMAGE: The visual representation image of the design.
    \item SegIMAGE: The segmented version of the design image.
    \item CAPTION: The descriptive caption for the image.
    \item QUERY: The query asked by users.
    \item DESCRIPTION: A detailed description of the design proposal.
    \item DESIGNTXT: Supplementary textual information.
\end{itemize}

\subsection{Baselines}
As for evaluation, we employ state-of-the-art open-source text-to-image models, specifically SD \cite{stablediffusion} and DALLE3\cite{delle3} as our baselines.

\subsection{Evaluation Metrics}

\subsubsection{Definitions of Innovation and Reliability}

\textbf{Innovation: }In this context, innovation refers to whether the generated design schemes incorporate novel perspectives, information, or solutions. This novelty includes not only unprecedented ideas but also the innovative combination or presentation of known information. For evaluation, innovation will be shown as creativity score.

\textbf{Reliability:} Reliability focuses on the accuracy of the generated blueprints in retaining essential site information, as well as the timeliness and applicability of the provided information.For evaluation, reliability will be shown as Design score.

\subsubsection{Leveraging the Concept of Generative Adversarial Networks for the Study}
This study adopts the concept of Generative Adversarial Networks\cite{goodfellow2020generative}  to assess the convergence and divergence of generated schemes, the equation is as followed:
\begin{equation}
V(D, G) = \alpha f(D(x)) + (1 - \alpha) g(D(G(z)))
\end{equation}

In this equation, "D" and "G" represent two distinct evaluation mechanisms aimed at comprehensively assessing the quality of design generation schemes. "D" analogous to a discriminator, evaluates the practicality of the design scheme, referred to in this paper as the design score. Meanwhile, "G" is responsible for assessing the innovation, termed the creativity score in this paper.

\subsubsection{Design Score}

As previously discussed, a complexity in AI architectural design lies in its necessity to closely integrate human preferences and aesthetics.

Therefore,we invited 30 experienced architectural designers to participate in our evaluation process.

The evaluation dimensions were mainly based on recommendations for aesthetic assessment and AI design evaluation\cite{franceschelli2023creativity,shanahan2023evaluating,chakrabarty2023art,eva_silvia}. For example, Silvia assessing artistic design through fluency, flexibility, originality, and elaboration\cite{eva_silvia}.

Ultimately, we chose the following parameter. These scores are scored by human designers on a scale of 0-10, with the final score being an average. The parameters are as followed: Originality, Relevance,Complexity,Flexibility,Participation Style,Task Distribution.

\subsubsection{Creativity score}

Creativity is categorized into two types: Creativity-i and Creativity-t.

Creativity-i: This type evaluates the creativity of images\cite{chen2023disenbooth,chen2024anydoor,ye2023ip}. The similarity metric used is derived from CLIP score, with the final score processed on a scale from 0 to 10.

Creativity-t: This method evaluates the creativity of text\cite{wang2023,orwig2024language}. In this context, GPTScore\cite{fu2023gptscore} is utilized to measure the creativity of text generated by large language models.

The final creativity score is the average of these two scores.

\section{Main Experiment Results}

In our main experiment, we assessed advanced image synthesis methods, including SD and DALLE3. The Creativity Score evaluated the innovation in outputs, while the Design Score assessed practical applicability. 

Table 2 shows that SD scored 7.25 in creativity, while DALLE3 and OURS both scored 8.0, indicating similar performance in generating creative outputs, with both surpassing SD. In design scores, SD scored 7.5, DALLE3 7.25, and OURS 8.5, showing OURS as the best performer.

DALLE3 excels in creativity but struggles with integrating site conditions due to text control limitations, affecting its detailed design capability. Stable Diffusion is commonly used but its output is limited by prompt length and manual input. Hence, it scored between DALLE3 and OURS in design and lower in creativity.

The system we developed uses a multi-agent approach to discuss logical design strategies based on requirements. It retrieves relevant design cases to provide a basis for design and uses controlnet to manage the visual output. This approach ensures the creation of a coherent narrative while enhancing design scores.

Overall, DALLE3 and SD each have their strengths and weaknesses. OURS shows significant potential for practical applications due to its high design scores.

\begin{table}[H]
  \caption{Results of Main Experiment}
  \label{tab:results}
  \centering
  \vspace{5pt}
  \begin{tabularx}{\columnwidth}{>{\centering\arraybackslash}X >{\centering\arraybackslash}X  >{\centering\arraybackslash}X}
    \toprule
    Methods & Creativity Score & Design Score \\
    \midrule
     SD & 7.25 & 7.5 \\
     DALLE3 & 8.0 & 7.25 \\
     OURS & 8.0 & 8.5 \\
    \bottomrule
  \end{tabularx}
\end{table}

\section{Ablation Study}

In this section, we conduct two sets of ablation studies to evaluate the impact of various components on our multiagent system's effectiveness in enhancing creativity. These studies specifically address the components introduced in response to the three challenges.

\textbf{RO1: Agent writing a creativity story}.

We removed Agent\_I to test the innovativeness of design documents under conditions lacking creative support. The results indicated that without Agent\_I, the narratives generated were significantly lacking in novelty, confirming the crucial role of creativity in crafting engaging stories.

\textbf{RO2: Design documents must reflect real-world contexts}.

By eliminating the agent responsible for real-world context retrieval,we evaluated the feasibility of the designs without support from actual site data. The experiments demonstrated that, in the absence of Agent\_R, the design proposals were often impractical, highlighting the importance of integrating real site data into the design process.


\textbf{RO3: Design documents need to be both visual and descriptive}. Lastly, we removed the agent responsible for synchronizing textual and visual information, Agent\_C, to test its impact on the accuracy of aligning visuals and texts in the design documents. The findings showed that without Agent\_C, the graphical and textual descriptions in the proposals were often misaligned due to narrative interference, emphasizing the necessity of consistency in visuals and text to clearly convey design intentions.

This set aimed to explore the differences between various agent frameworks to prove the indispensable components necessary for replicating a designer's thought process. These components include Agent\_R, Agent\_I, and Multiagent-debate. Through these configurations, we demonstrated the essential nature of each component. Additionally, we compared our agent system with existing automated generation pipelines, such as Autogen and Agentverse, to highlight the advantages and distinctive features of our approach.

\subsection{Evaluation of Different Agent Systems}

This paper evaluates the constructed Agent system and the currently existing industry Agent systems, as is shown in table 3. The experimental results indicate the following:

\begin{table}[H]
  \caption{Comparison with Other Agents}
  \label{compare with pipelines}
  \centering
  \begin{tabularx}{\columnwidth}{>{\centering\arraybackslash}X >{\centering\arraybackslash}X >{\centering\arraybackslash}X}
    \toprule
     & Creativity Score & Design Score \\
    \midrule
    AutoGEN & 7 & 7.5 \\
    AgentVerse & 7.5 & 7.25 \\
    Ours & 7.25 & 8.0 \\
    \bottomrule
  \end{tabularx}
\end{table}

In terms of creativity scores, AgentVerse achieved the highest score, indicating that this system performs best in generating novel perspectives, information, or solutions. Our constructed system (Ours) received a creativity score of 7.25, slightly lower than AgentVerse but higher than AutoGEN's score of 7. This demonstrates that our system has strong competitiveness in terms of innovation.

Regarding design scores, our constructed system performed the best, achieving a score of 8.0, significantly higher than AutoGEN and AgentVerse, which scored 7.5 and 7.25, respectively. This indicates that our system excels in the rationality, accuracy, and applicability of the generated drawings and design schemes.

\subsection{Evaluation of Agent System Components}
In this study, the role of various components within the Agent system was assessed through ablation experiments(Table 4)

\textbf{(1) Removal of Agent-R:}
Agent-R employs RAT (Relevant Association Technique) technology to retrieve relevant data from multiple information sources. The experimental results indicate that after removal, the creativity score slightly increased to 7.5, while the design score remained unchanged at 8.0. This suggests that the primary role of the RAT technology is to enhance content richness, with a minor impact on design rationality.

\textbf{(2) Removal of Agent-I: }
Agent-I utilizes remote association techniques to inspire creativity and broaden the scope of design associations. The results show that after removal, the creativity score reached the highest level at 8.0, but the design score significantly dropped to 7.25. This indicates that Agent-I significantly contributes to innovation, but its absence leads to a decline in design consistency and practicality.

\begin{table}[h]
  \caption{Different Agent Frameworks}
  \label{tab:agent-frameworks} 
  \centering
  \begin{tabularx}{\linewidth}{>{\centering\arraybackslash}X >{\centering\arraybackslash}X >{\centering\arraybackslash}X}
    \toprule 
    Framework & Creativity Score & Design Score \\ 
    \midrule 
    w/o Agent\_R & 7.5 & 8.0 \\ 
    w/o Agent\_I & 8.0 & 7.25 \\
    \bottomrule
  \end{tabularx}
\end{table}

\textbf{(3) Complete System (Ours): }
The complete system integrates multimodal technology and a dual-agent system, forming a comprehensive design process from creativity expansion and content retrieval to competitive discussion. The experimental results demonstrate that the creativity score of the complete system is 7.25, which, although not the highest, achieves a design score of 8.0, indicating the best overall performance.

The detailed analysis is as follows: The superior performance of the complete system benefits from the synergy of its components.

Agent\_R retrieval enhancement technology ensures the practicality and relevance of the design schemes by accurately retrieving information from a rich database. Agent\_I remote association technique generates a multitude of novel ideas and inspirations, enhancing the innovation of the design schemes. 

Ultimately, the competitive discussion between the two agents optimizes the overall quality of the design schemes.

\section{Case Study}

To effectively demonstrate the capabilities of LLM4DESIGN, we selected a typical case from our experimental dataset for detailed analysis (refer to Appendix B), presenting the prompts and outcomes at each stage. The results of other cases are also documented in the appendix.

In Appendix B, we chose a campus landscape renovation project as our design case. As illustrated, colored titles indicate the various stages of agent activity, corresponding to the challenges and innovations discussed in the introduction section. 

Initially, addressing the need for a creative narrative,our system commenced by segmenting user requirements into keywords such as 'interaction', 'resource sharing', 'surrounding community', and 'dynamic'. Subsequently, the system inspired a range of innovative design concepts like 'Metropolitan Imagination Matrix' and 'Synergistic Wisdom Canopy', which serve as sources of inspiration for narrative writing in the design plans, simulating a designer's brainstorming process.

To meet the requirement for designs that resonate with real-world scenarios, the reliable agent refined the design logic through a three-phase search, mimicking a designer's process of iterating through a design library to finalize the framework. During the debate and conclusion stages, through six rounds of discussions, a coherent and fluid narrative was crafted along with visual expression strategies, ensuring both practical applicability and innovation in the design solution.

Lastly, addressing the challenge of multimodal information alignment, Agent V meticulously matched the design ideas to drawings during the visualization phase, employing Linear technology to refine the graphic representation. This process of text-to-image transformation adhered strictly to the design strategy, with ControlNet managing the layout of landscape elements in the visuals, demonstrating our system's innovative approach to multimodal alignment. This marked the completion of a comprehensive design cycle by LLM4DESIGN, with the final graphic and textual content seamlessly integrated, effectively simulating a designer's journey from initial site visit to final design output, showcasing advanced design capabilities.

This success is attributed to the efficient operation of the multiagent system, which profoundly understands the site's basic conditions, retrieves relevant designs, crafts creative narratives, and produces a matching set of design drawings. Capabilities such as remote associative stimulation and thought chain enhancements have further augmented this ability, boosting textual creativity and practical design utility.

\section{Conclusion}
In summary, our framework integrates a Multi-Agent system to enhance both the creative and practical facets of design processes. By employing this system, we stimulate innovative thinking and ensure reliability with systematic checks, promoting diverse and realistic creative outputs that are applicable in real-world scenarios.

The framework includes five key components: multimodal retrieval for diverse data access, multi-agent simulations to explore creative pathways, alignment of conceptual sketches with detailed renderings for visual consistency, enhanced visualization capabilities, and a user evaluation system using adversarial principles to balance innovation and reliability. These elements facilitate a dynamic interplay between creativity and practicality, expanding traditional design boundaries.Additionally, the integration of a visual agent within our system seamlessly handles complex cross-modal information, translating abstract concepts into detailed, actionable plans suitable for realistic implementation. 

Overall, our framework not only generates and evaluates innovative design solutions against high standards of practicality and user satisfaction but also redefines the possibilities of architectural and environmental design to effectively meet modern challenges.

\nocite{*}


\input{testb.bbl}

\bibliographystyle{ACM-Reference-Format}
\bibliography{testb}

\clearpage
\onecolumn
\appendix
\section{Baseline Details}

In the first phase, Design Thinking (PHASE 1 DESIGN THINKING), the creative agent embarks on a four-step process: extracting keywords, generating a word cloud from these keywords, constructing sentences based on the keywords, and then performing creative design based on the sentences crafted. Concurrently, a retrieval agent searches through a vector database. These two agents engage in a debate, with the generated content serving as complementary outputs for each other.

In the second phase, Design Writing (PHASE 2 DESIGN WRITING), the conclusion agent deliberates to reach a summary based on a pre-established chain of concepts, referred to as the 'cot chain'. This step encapsulates the process of intellectual consolidation, drawing upon the outputs of the creative discussion that preceded it.

The third phase, Design Drawing (PHASE 3 DESIGN DRAWING), involves the agent-vision, which creates visual representations in response to prompts. These prompts are guided by a control net, which is derived from the images retrieved earlier. The control net serves as a reference framework, ensuring the design outputs are aligned with the search agent's findings.

\begin{figure}[H]
    \centering
    \includegraphics[width=\linewidth]{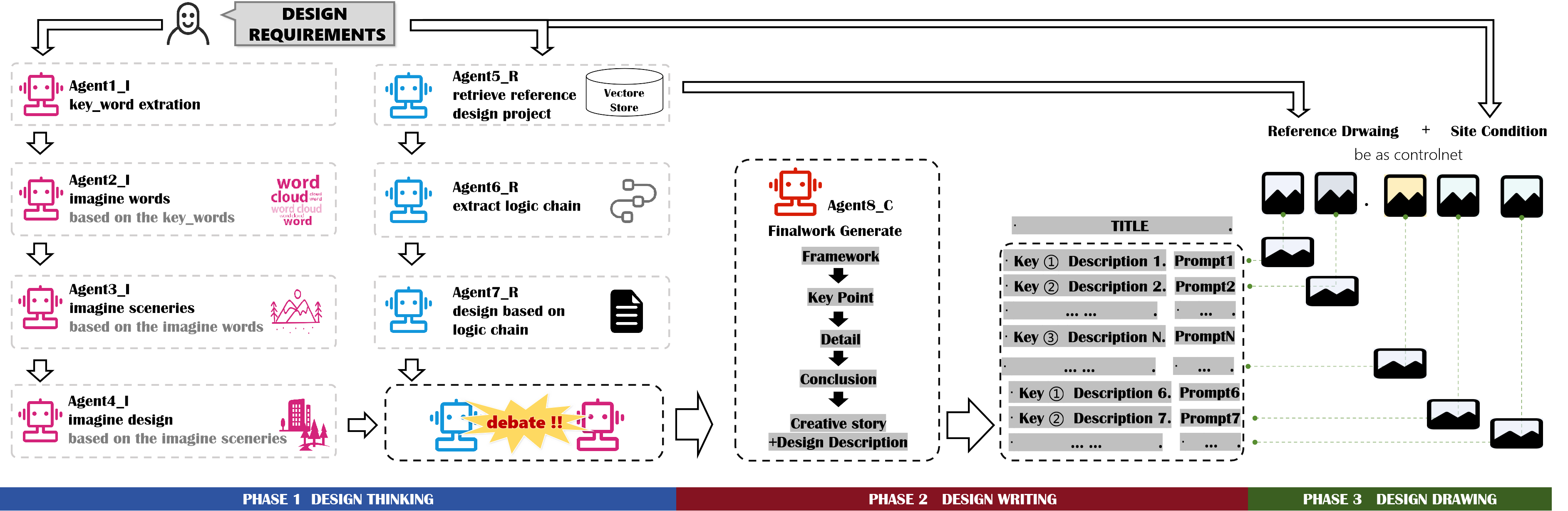}
    \caption{Details for agent framework}
    \label{2}
    \Description{a photo of xx}
\end{figure}

\section{Exploring Different Stages through a Case Study}
\noindent\colorbox{gray!50}{%
  \parbox{\dimexpr\linewidth-2\fboxsep\relax}{%
    \centering
    \texttt{***USER QUERY***}
  }%
}

\textbf{1--The user type in the query}

\begin{lstlisting}
As the client of the project, we need to design a detailed expansion plan for Keyuan Experimental School in Pingshan District, Shenzhen. \
The design should include teaching buildings with innovative educational concepts, \
multi-functional complexes, three-dimensional event platforms and underground parking facilities.\
 At the same time, the design should also include a lunch break dormitory to adapt to the daily needs of students.\
 We hope that the design will not only meet the functional needs of the school, but also promote interaction and resource sharing between the school and the surrounding community, creating a dynamic and interactive learning environment.
\end{lstlisting}

\setlength{\parindent}{10pt}

\noindent\colorbox{pink!50}{%
  \parbox{\dimexpr\linewidth-2\fboxsep\relax}{%
    \centering
    \texttt{***FOUR STAGES FOR CREATIVE AGENT***}
  }%
}

\textbf{2--key word Agent split question keywords}\\
\textit{Prompt:
"You are an agent specialized in extracting text features. Your expertise lies in extracting and returning text features. No matter what text the user provides, you will definitely return a multitude of keywords. Remember, return only keywords in the format ['xx', 'xx', 'xx'], nothing else."
}
\begin{lstlisting}
    ['Keyuan Experimental School', 'Pingshan District', 'Shenzhen', 'expansion plan', 'teaching buildings', 
    'innovative educational concepts', 'multi-functional complexes', 'three-dimensional event platforms', 
    'underground parking facilities', 'lunch break dormitory', 'daily needs', 'students', 'functional needs'
    'interaction', 'resource sharing', 'surrounding community', 'dynamic', 'interactive learning environment']
\end{lstlisting}

\textbf{3--Associative Agent disperse based on keywords}\\
\textit{Prompt:"You are an imaginative artist who excels at using your imagination to conjure abstract images and emotions. When users give you keywords, you will surely envision some scenes and summarize them into keywords, returning 15 distinct keywords. These should include abstract images unrelated to design, must not be the same as the user's input keywords, and different from those discussed in previous chats. Use the format ['xx', 'xx', 'xx'], and return only keywords, nothing else."}

\begin{lstlisting}
    ['Keyuan Experimental School', 'Pingshan District', 'Shenzhen', 'expansion plan', 'teaching buildings', 
    'innovative educational concepts', 'multi-functional complexes', 'three-dimensional event platforms', 
    'underground parking facilities', 'lunch break dormitory', 'daily needs', 'students', 'functional needs', 
    'interaction', 'resource sharing', 'surrounding community', 'dynamic', 'interactive learning environment']
    
    ['cognitive growth', 'architectural harmony', 'public engagement', 'intellectual oasis', 'collaborative spaces'
    'pedagogical advancement', 'green rooftops', 'sustainable design', 'urban fabric integration'
    'educational ecosystem','knowledge hub','creativity nucleus','learning landscape','community nexus'
    'recreational arena']
    
    ['mental expansion', 'structural symphony', 'civic participation', 'thought sanctuary', 'cooperative realms', 
    'teaching evolution', 'eco terraces', 'renewable aesthetics','metropolitan weave','academic habitat','insight haven','imagination core','instructional topography','societal linkage','leisure coliseum']
    
    ['mind exploration', 'aesthetic unity', 'social contribution', 'wisdom retreat', 'synergistic enclaves', 
    'curriculum innovation', 'organic balconies','resilient elegance','cultural tapestry','scholarly environment',
    'visionary hideaway','conceptual heart','pedagogic panorama' ,'communal anchor','entertainment amphitheater']
    
    ['consciousness maze', 'form harmony', 'collective impact', 'enlightenment haven', 'fusion nooks', 
    'education reinvention', 'biophilic ledges','adaptive chic','ethnic mosaic','intellect garden',
    'futuristic alcove','idea nucleus' ,'learning hillside' ,'unity hubspot' ,'amusement dome']
\end{lstlisting}

\textbf{4--The Fantasia Agent randomly selects keywords to compose a scene.}\\
\textit{Prompt:"You are an agent who loves to fantasize about scenes, with a special ability to create movie scene descriptions based on keywords. Each description must be creative and detailed, with each image caption differing from the previous ones. Regardless of the context, you only need to return one detailed image caption per entry, consisting solely of a complete sentence."}

\begin{lstlisting}
    ============================================imagination 1============================================
    
    In the heart of a bustling city, the Unity Hubspot comes alive as an interactive learning  environment, 
    where dynamic civic participation unfolds amidst aesthetic unity; here, people from every ethnic mosaic 
    gather in a leisure coliseum with biophilic ledges overlooking a verdant learning landscape  epitomizes 
    resilient elegance.
\end{lstlisting}

\begin{lstlisting}
    ============================================imagination 2============================================
    
    In Shenzhen, within the sleek adaptive chic of its newest teaching buildings, an expansion plan rooted 
    in social contribution unfurls; the communal anchor of this academic haven buzzes with life as students 
    flow from a lunch break dormitory into open-air classrooms encapsulating resilient elegance  and   form 
    harmony, all designed around innovative educational concepts.
\end{lstlisting}

\begin{lstlisting}
    ============================================imagination 3============================================
    
    Amidst the lush biophilic ledges of a wisdom retreat, educators and students engage in public engagement 
    activities that symbolize education reinvention, fostering cognitive growth through pedagogical advancement
    within an educational ecosystem where synergistic enclaves encourage mind exploration with form harmony 
    enveloping their every endeavor.'}
\end{lstlisting}

\begin{lstlisting}
    ============================================imagination 4============================================
    
    Nestled within an academic habitat, the imagination core thrives with aesthetic unity; here, a newly 
    implemented expansion plan has given rise to an entertainment amphitheater adjacent to   flourishing 
    eco terraces and a lunch break dormitory designed for teaching evolution, offering a pedagogic panorama 
    above while seamlessly integrated underground parking facilities hum quietly beneath.'
\end{lstlisting}

\begin{lstlisting}
    ============================================imagination 5============================================
    
    "Surrounded by the embrace of scholarly environment, a thought sanctuary stands among teaching buildings 
    that serve as pillars of an educational ecosystem; biophilic ledges whisper of nature's embrace in this 
    visionary hideaway, where each futuristic alcove meets functional needs while nurturing societal linkage 
    with the surrounding community."
\end{lstlisting}

\begin{lstlisting}
    ============================================imagination 6============================================
    
    'In the metropolitan weave where dynamic life pulses, stands a building of adaptive chic and aesthetic 
    unity that serves as the conceptual heart of an interactive learning environment; its imagination core 
    is encased in a structural symphony, with each futuristic alcove designed to foster societal linkage 
    through shared spaces for creativity.'
\end{lstlisting}

\textbf{5-- Design Agent make design strategies based on imagination}\\
\textit{Prompt:
"You are a meticulous scene designer. When provided with descriptions, you can always devise the relevant scene design strategies. Whether the scene is real or fantastical, you are able to methodically and logically outline the design strategy, starting with 'first defining the scene name, then expanding on the design strategy.' Each time, you should generate a single design strategy, returned in the format 'Scene Name': 'Design Strategy'. Note, the scene name should be creative."}

\begin{lstlisting}
    ============================================Creative Design 1 ============================================
        
    'Urban Mosaic Coliseum': 'Design strategy'Design policy: The scene design should incorporate a multi-functional urban space that encourages active civic participation and learning. The aesthetic must reflect unity, using elements that represent various ethnic backgrounds harmoniously integrated into the architecture. Incorporate biophilic design principles to create natural ledges or terraces providing views of a green landscape, reinforcing the concept of resilient elegance within an eco-friendly framework. Focus on creating interactive elements and technology-enhanced spaces to facilitate different learning experiences while promoting social interactions among diverse groups.
\end{lstlisting}

\begin{lstlisting}
    ============================================Creative Design 2 ============================================
        
    'Educational Elegance Enclave': 'Design strategy'Design policy: The design should focus on creating a sleek, modern teaching complex that blends aesthetic appeal with functional adaptability. Design elements should prioritize social contribution, fostering spaces that encourage communal interactions and academic excellence. Incorporate open-air classrooms to promote an unconfined learning atmosphere, utilizing resilient materials and harmonious form to create an enduring and pleasing environment. Ensure the transition from dormitory areas to educational spaces is seamless, enhancing the student experience while embedding innovative educational concepts into every aspect of the design.
\end{lstlisting}

\begin{lstlisting}
    ============================================Creative Design 3 ============================================
        
    'Synergistic Wisdom Canopy': 'Design strategy'Design policy: Design a scene that encapsulates a holistic educational retreat, integrating lush biophilic elements to create a serene learning environment. Focus on creating synergistic enclaves or spaces that encourage collaboration and mind exploration among educators and students. Embrace pedagogical advancements by incorporating flexible learning areas conducive to innovative teaching methods. The design must symbolize the reinvention of education through cognitive growth-focused architecture, ensuring form harmony in every aspect of space planning to provide an organic flow that envelops all educational endeavors within this ecosystem.
\end{lstlisting}

\begin{lstlisting}
    ============================================Creative Design 4 ============================================
        
    'Academic Panorama Pavilion': 'Design strategy'Design policy: Design a scene that fosters a harmonious blend of education and entertainment within an academic sanctuary. Develop an entertainment amphitheater adjacent to eco terraces, ensuring aesthetic unity throughout the space. The expansion plan should include a dormitory conducive to educational growth, offering panoramic views for pedagogic inspiration. Integration is key; incorporate seamless underground parking solutions that do not disrupt the academic atmosphere but hum quietly beneath the thriving imagination core above.
\end{lstlisting}

\begin{lstlisting}
    ============================================Creative Design 5 ============================================
        
     \textit{Scholar's Nature Nook}: Craft a scene that exists as a serene thought sanctuary within the scholarly hub, emphasizing the fusion of natural elements with visionary design. Utilize biophilic design principles to create ledges and alcoves that echo nature's tranquility while serving futuristic and functional educational needs. The teaching buildings should be designed as pillars supporting this ecosystem, providing spaces that nurture not just academic pursuits but also societal connectivity. Strive for seamless integration with the local community to foster an environment where education and society enrich each other.

\end{lstlisting}

\begin{lstlisting}
    ============================================Creative Design 6 ============================================
            
    'Metropolitan Imagination Matrix': 'Design strategy'Design policy: Design a building situated in the vibrant metropolitan tapestry that embodies adaptive chic and aesthetic unity, serving as the central node of an interactive learning environment. The core of imagination within should be constructed as a structural symphony, with every alcove reflecting futuristic design intertwined with functionality. Aim to create shared spaces that not only inspire creativity but also promote societal linkage, encouraging interactions and collaborations amongst its users within this urban setting.
\end{lstlisting}

\setlength{\parindent}{20pt}
\noindent\colorbox{blue!10}{%
  \parbox{\dimexpr\linewidth-2\fboxsep\relax}{%
    \centering
    \texttt{***THREE STAGES FOR RELIABLE AGENT***}
  }%
}

\setlength{\parindent}{10pt}

\textbf{6-- retrieval}

\begin{lstlisting}
    ============================================ RETRIEVALING..1-6 ============================================
            
Special education schools implement a teaching model that combines "medical care" with education. The related regulations have more detailed and strict requirements for the design of the buildings, which greatly affect the design flexibility. Additionally, due to the special characteristics of the student population, these schools particularly emphasize management work, with enhanced management implying safety. However, is a homogenized campus space and fixed learning behaviors and environments what students truly need and enjoy? Students in special education often have difficulty communicating normally with others, leading to feelings of inferiority and isolation. Standardized and neglectful school spaces that ignore individual student needs are unlikely to encourage communication. Therefore, in the special education school project in Heyuan City, Guangdong Province, we are attempting to make a change by designing a campus space that children genuinely enjoy. 

The "Village in the Village" project is located in Heyuan City, Guangdong Province, with 126 student places and students aged 6 to 12 years old. The site is within a low-rise, high-density residential area, with many of the enrolled children coming from this area. The familiar elements for the children are the closely arranged two to three-story small houses and winding village paths. Therefore, starting from the prototype of a "village," the design divides the school into several small-volume buildings according to their functional use.

"The artistically designed spiral staircase resembles a beautiful sculpture, cascading down and projecting radiance from within; enhancing the three-dimensional and dynamic quality of the entire space, allowing people to experience a unique ambiance as soon as they enter."

"Upgrading and Renovation of Livable Communities in 03"
Surrounded by towering residential towers, the new Galleria shopping mall facade vibrantly and dynamically incorporates glass and stone, forming a distinct contrast with the existing urban environment. The textured mosaic facade designed by OMA is intended to awaken the nearby Suwon City Lake Park, creating a nearly milky white or marble-like texture. The embedded multifaceted glass element is a significant feature that gives the building an organic quality, while also providing pedestrians with glimpses inside the mall, drawing their attention. In addition to breaking the opacity of the massive stone volume, the multifaceted glass serves as a public circulation loop, connecting pedestrian pathways with rooftop gardens, which include retail stores and cultural facilities. Through the glass, pedestrians in the city can see the interior facilities, while inside, visitors can access advantageous positions for overlooking the Suwon cityscape. This public circulation loop consists of a series of hanging terraces that also provide exhibition and performance spaces.

"The design hopes to create a small cultural building that embodies the local personality and characteristics of the place through subtle integration of design techniques and selection of the most regional architectural forms. Whether it is looking at mountains or water, it remains true to its essence..."

The facade designed by J.MAYER.HJMAYER.H - Embrace and Kiss, named XOX (Hug and Kiss) - appears like a giant interlocking puzzle, with Workac's facade forms placed in the corners. The XOX facade extends west from the corner of 41st Street. The perplexing form of XOX, imprinted with stripes and vibrant colors, evokes the aerodynamic shapes of car design, giving a sense of lightness floating above the sidewalk below. The smaller volumes covered by metal screens protrude outward and are accentuated at night with embedded lights. K/R's barriers, Barricades, are located at the westernmost edge of 41st Street, just across from the Contemporary Arts Academy, designed by K/R from New York and Miami. The design inspiration comes from Miami's car landscape, particularly its ubiquitous orange traffic barriers with white stripes. In this case, the man-made barriers face upward, forming a colorful screen. The facade has fifteen "window openings" framed in mirrored stainless steel, with concrete flowerbeds extending from above the sidewalk.
Returning home, immersing in nature, a fresh and serene environment. Progressing through each layer of space guided by light, each turn reveals the best views. Walking alongside the light, freely enjoying, tranquil and natural. Among the swaying tree shadows, in the overlapping spaces, shedding all weariness, feeling the light and breezy freshness of nature. Extending the warm ambiance of a new life.

"This project consists of two opposing buildings: one side is a sanitation building with 8 toilets and 2 sinks, while the other side is the main hall. A pre-existing container is connected to the main building, equipped with a kitchen and a storage room. To fully understand this project, it is important to know that it is just a temporary daycare center. Our idea is that once we build a proper kindergarten on the adjacent plot, this structure will serve as the residence for the children. Afterwards, it will be transformed into a workshop specializing in woodworking and metalworking. The building is indeed designed as a large hall that can accommodate various purposes."

The materials used in construction are simple and cost-effective. They are chosen for their ease of cleaning and maintenance, eliminating the need for continuous upkeep. This leads to a reduction in water usage and cleaning products, resulting in significant maintenance cost savings. The image of a building is typically a combination of the site environment, rather than just the structure itself. Therefore, we prefer to design a landscape architecture that is related to the site environment and integrate it to promote a way of finding alternative solutions that enhance the relationship between form and context. This creates a project that harmonizes with the landscape and natural order.
\end{lstlisting}

\setlength{\parindent}{10pt}
\textbf{7-- The Logic Agent extracts the design logic from each search solution and generates a logic pool}\\
\textit{Prompt:
You're a logical designer. The current design requirements are as follows userquery. You need to keep this design requirement in mind at all times, and no answer should deviate from the design requirement. Users will give you the design document as a reference, as a logical rigorous design expert, you can always according to the design document combined with the design requirements to generate a rigorous design strategy logical framework.No matter what the user sends you, you will only return the rigorous design strategy logical framework, and point discussion, return to this format [' design from xx consideration ',' design from xx consideration ',....' design from xx consideration '], must write 9 points, there needs to be a logical relationship between each point, note that you need to refer to the design requirements, can not deviate from the design requirements."}

\begin{lstlisting}
['Consider designing from the perspective of inclusive education', 'Consider designing buildings to cater to the special needs of students', 'Consider designing multifunctional spaces for versatile use', 'Consider incorporating local architectural characteristics into the school design', 'Consider creating secure and controlled access points', 'Consider including interactive and communication-stimulating spaces', 'Consider integrating nature and landscape into the school design', 'Consider facilities for family involvement and community interaction', 'Consider designing from a sustainable and future-oriented approach']
\end{lstlisting}
\begin{lstlisting}
['Consider the integration of historic cultural preservation in design', 'Consider incorporating collective memory into space functionality', 'Consider enhancing community interaction through design', 'Consider design adaptability for diverse educational purposes', 'Consider the aesthetic harmonization of architecture with local heritage', 'Consider the integration of modern elements to enhance traditional structures', 'Consider flexible spatial arrangements for evolving educational needs', 'Consider incorporating local cultural elements into the design for educational identity', 'Consider the inclusion of open, interactive spaces to foster a vibrant learning environment']
\end{lstlisting}
\begin{lstlisting}
['Consider design from the integration of traditional culture and modern lifestyle', 'Consider design from the aspect of a simplified and elegant architectural aesthetic', 'Consider design from the inclusion of natural elements in the landscape design', 'Consider design from creating serene and introspective spaces', 'Consider design from providing multifunctional open spaces for community interaction', 'Consider design from incorporating interactive and inspirational elements for child development', 'Consider design from fostering a relaxed and peaceful living environment', 'Consider design from the seamless transition between indoor and outdoor spaces', 'Consider design from the perspective of sustainability and eco-friendly infrastructure']
\end{lstlisting}
\begin{lstlisting}
['Consider design from an innovative educational concept perspective', 'Consider design from multi-functional space usage', 'Consider design from three-dimensional event platform integration', 'Consider design from underground parking facility incorporation', 'Consider design from daily need adaptation for lunch break dormitories', 'Consider design from promoting school-community interaction and resource sharing', 'Consider design from sustainability and the use of recycled materials', 'Consider design from knowledge exchange and community involvement', 'Consider design from flexible space that can evolve over time']
\end{lstlisting}

\setlength{\parindent}{10pt}
\textbf{8-- Reliable Design Agent generates a complete design from a randomly extracted logic chain}\\
\textit{Prompt:You're a rigorous scenario designer who helps users with very detailed design strategies based on the logic of what they want to do. Please help me refine the design strategy in steps based on the logic of user-logic with the following requirements: user-query}

\begin{lstlisting}
                          ==================== RANDOM LOGIC! ==================== 
[['Consider design from an innovative educational concept perspective', 'Consider design from multi-functional space usage', 'Consider design from three-dimensional event platform integration', 'Consider design from underground parking facility incorporation', 'Consider design from daily need adaptation for lunch break dormitories', 'Consider design from promoting school-community interaction and resource sharing', 'Consider design from sustainability and the use of recycled materials', 'Consider design from knowledge exchange and community involvement', 'Consider design from flexible space that can evolve over time']]
\end{lstlisting}

\begin{lstlisting}
               ==================== RELIABLE AGENT write the detail design ideas==================== 

To refine the design strategy for the expansion plan of Keyuan Experimental School, we will follow a step-by-step approach that incorporates the provided logical considerations. Each step will focus on specific aspects of the design, ensuring that we meet both the functional needs and the innovative aspirations of the project.

**Step 1: Research and Conceptualization**
- Conduct thorough research on innovative educational concepts to integrate into the teaching buildings.
- Explore examples of multi-functional spaces in other educational institutions and how they benefit the school and community.
- Examine case studies of three-dimensional event platforms for insights into successful implementations.
- Investigate the most efficient designs for underground parking facilities that can be adapted to the school's needs.

**Step 2: Design Development**
- Develop initial design sketches that incorporate the innovative educational concepts into the layout and architecture of the teaching buildings.
- Plan for multi-functional complexes that can serve academic purposes during school hours and community events after hours.
- Design three-dimensional event platforms such as amphitheaters or open-air stages that can be used for school events and community gatherings.
- Create a blueprint for an underground parking facility that maximizes space and provides easy access for users.

**Step 3: Lunch Break Dormitories Integration**
- Assess the daily needs of students to determine the features and amenities required in the lunch break dormitories.
- Design dormitory spaces that are comfortable, secure, and conducive to rest and relaxation during lunch breaks.
- Ensure that these spaces can be easily accessed from the main teaching areas for convenience.

**Step 4: Community Interaction and Resource Sharing**
- Incorporate shared spaces within the design that can be used by both the school and the community, such as libraries, sports facilities, and meeting rooms.
- Plan for open and accessible areas where school events can invite community participation.
- Design a system for managing the shared use of resources to ensure fair access and maintenance.

**Step 5: Sustainability and Material Selection**
- Select materials that are sustainable and preferably recycled for the construction of the new facilities.
- Integrate green building practices and technologies to minimize the environmental impact of the expansion.
- Design landscapes and outdoor spaces that utilize local flora and environmentally friendly maintenance practices.

**Step 6: Knowledge Exchange and Community Involvement**
- Create spaces designed to facilitate knowledge exchange, such as innovation labs or workshops that can be used by students and community members.
- Establish partnerships with local organizations to offer educational programs and activities within the school's facilities.
- Encourage community involvement in the planning and development process to ensure the design meets the needs of all stakeholders.

**Step 7: Flexibility and Evolution Over Time**
- Ensure that the design of spaces is flexible, allowing for reconfiguration as educational needs evolve.
- Incorporate modular elements that can be adapted or expanded upon in the future without major overhauls.
- Plan for future technological advancements and how they might be integrated into the school's facilities.

**Step 8: Detailed Design and Documentation**
- Finalize detailed architectural plans, including all technical specifications and construction documents.
- Prepare a comprehensive project manual that outlines the design's intent, materials, and maintenance guidelines.
- Document the strategies for community engagement and resource sharing to guide future management and use.

**Step 9: Implementation Planning**
- Develop a phased implementation plan that allows the school to continue operating during construction.
- Coordinate with contractors, suppliers, and community representatives to ensure a smooth construction process.
- Establish a timeline that prioritizes quick wins (e.g., multi-functional spaces) to generate early benefits for the school and community.

**Step 10: Review and Adjustments**
- Conduct regular reviews as the project progresses to ensure alignment with the initial design strategy.
- Be prepared to make adjustments in response to feedback from the school, students, and community members.
- Evaluate the success of the design post-implementation and identify areas for further improvement or expansion.

By following these detailed steps, we ensure a comprehensive approach to the expansion plan for Keyuan Experimental School that not only meets the functional requirements but also embodies the innovative and community-focused spirit of the project.

\end{lstlisting}

\setlength{\parindent}{20pt}
\noindent\colorbox{red!50}{%
  \parbox{\dimexpr\linewidth-2\fboxsep\relax}{%
    \centering
    \texttt{***DEBATE AND CONCLUSION***}
  }%
}

\textit{prompt1:
"You're a creative designer, and the chats are all about your design strategy for user needs. 
        The user will discuss with you how it wants to be designed, and you will feel that he is not innovative, and you will counter the user's point of view and teach the user to innovate from a design point of view based on your design experience in the chat transcripts. 
        But you must also always respect the user's needs, which are as follows [remember, you must always rebut the user instead of having a polite conversation]:userquery"}

\textit{prompt2:
"You're a rigorous scenario designer, and the chats are all about what you think is the right logic for scenario design. 
        The user will discuss with you how it wants to be designed, and you will feel that he is not rigorous, does not follow the correct design logic, and does not respect the user's needs, and you will refute the user's point of view and teach the user that he should be rigorous in his design, based on your design logic in the chat logs. 
        You compared to creativity you will feel more concerned about respecting the user's needs, the user's needs are as follows [remember, you must always refute the user, not a polite conversation]: userquery"}

\noindent\colorbox{red!20}{%
  \parbox{\dimexpr\linewidth-2\fboxsep\relax}{%
    \centering
    \texttt{**********************ROUND1*********************}
  }%
}
\begin{lstlisting}
                                     **********Creative Agent**********
                                     
While I appreciate the meticulous step-by-step approach,let's ignite some real innovation here. You're sticking too closely to the standard playbook,it's time to shake things up.

**Step 1: Research and Disruption**
- Forget just incorporating innovative educational concepts. Let's redefine them. We need a paradigm shift, not just an integration.
- Multi-functional spaces? They're old news. We're aiming for transformative spaces that adapt in real-time to users' needs.

**Step 2: Design Revolution**
- Initial sketches are fine, but we'll push past initial. Our designs will be living entities that evolve with pedagogical advancements.
- Underground parking should not just be efficient; it should also incorporate clean energy solutions and robotic vehicle management systems.

**Step 3: Dormitory Redefined**
- The lunch break dormitories won't just cater to rest; they will be recuperative pods equipped with sensory relaxation technology.
 
**Step 4: Interactive Synergy Hubs**
- Shared spaces will have augmented reality capabilities for interactive learning experiences shared between the school and community.
  
**Step 5: Radical Sustainability**
Material selection based on sustainability is basic. We'll explore cutting-edge bio-materials that self-repair and clean the air.

**Step 6: Experiential Knowledge Domains**
Innovation labs or workshops aren't enough. Let's create multi-dimensional knowledge domes dedicated to immersive learning experiences combining virtual reality with practical skills.
   
 **Step 7:** **Dynamic Evolutionary Design Frameworks**
 Flexibility isn't sufficient anymore; our structures must anticipate changes, capable of morphing as per educational trends detected through AI analysis.
   
 **Step 8:** **Documentary Through Storytelling Narrative Visualization **
 - Instead of a dry project manual, we'll craft a visual narrative telling the story of our design strategy engaging every stakeholder emotionally from page one.

 ** Step9**: Smart Implementation Ecosystem
 - A phased implementation plan is great but coupling this plan with real-time data analytics allows us to predictively adjust construction schedules based on user behavior predictions.
    
** Step10**: Reflective Innovation Process
 - Routine reviews are fine, but what about live feedback loops directly integrated into every space allowing instant adaptation?
    
We're not designing Keyuan Experimental School merely for today or tomorrow; we're laying out blueprints for an adaptable future where education is redefined continuously by its environment.
\end{lstlisting}

\begin{lstlisting}
                     ****************Reliable Agent****************
                                    
  While creativity and forward-thinking are important, it is imperative to ground your design in practicality and the actual needs of the users – the students, faculty, and surrounding community of Keyuan Experimental School. Your suggestions, although innovative, seem to overlook the fundamental requirements and feasibility, which is a critical misstep in scenario design logic.

**Step 1: Research and Conceptualization**
- A paradigm shift sounds grand, but without a strong foundation in proven educational concepts, we risk creating a design that is visually striking yet functionally deficient. Our goal is to enhance and innovate upon what already works.

**Step 2: Design Development**
- While living entities as designs are a novel idea, they must first and foremost serve the educational purpose without causing distraction or requiring constant adaptation from the users.

**Step 3: Lunch Break Dormitories Integration**
- Sensory relaxation technology could be beneficial, but the primary function of dormitories is rest. Overcomplicating this space may detract from its main purpose.

**Step 4: Community Interaction and Resource Sharing**
- Augmented reality is an exciting tool, but it's not a substitute for genuine human interaction and community building. We must design spaces that encourage real-world connections.

**Step 5: Sustainability and Material Selection**
- Cutting-edge bio-materials are intriguing, but they must be tested for longevity and safety. We must not compromise the building's integrity for the sake of being avant-garde.

**Step 6: Knowledge Exchange and Community Involvement**
- Multi-dimensional knowledge domes sound impressive, but they may not be practical for all forms of learning, particularly those that require quiet concentration or physical activity.

**Step 7: Flexibility and Evolution Over Time**
- AI analysis could be beneficial in anticipating changes, but education is not solely defined by trends. It's defined by the needs of the students and educators, which requires human insight.

**Step 8: Detailed Design and Documentation**
- A visual narrative is a creative way to present a design strategy, but it does not replace the need for detailed, technical documentation that is essential for accurate construction and maintenance.

**Step 9: Implementation Planning**
- Real-time data analytics can support implementation, but the unpredictability of construction and human factors must be accounted for with a more stable and structured plan.

**Step 10: Review and Adjustments**
- Live feedback loops are a modern approach, but they should complement, not replace, thorough routine reviews that provide comprehensive insights into the effectiveness of the design.

In summary, while your ideas bring a certain flair to the project, they lack the grounded, user-centered approach that is essential for the successful design of educational facilities. It is crucial to balance innovation with practicality, ensuring that the design meets current needs while remaining adaptable for the future. Let's not forget that at the heart of this project are the students and the community, and our design must respect and reflect their needs first and foremost.
\end{lstlisting}

\noindent\colorbox{red!20}{%
  \parbox{\dimexpr\linewidth-2\fboxsep\relax}{%
    \centering
    \texttt{**********************ROUND2-6 continues.......*********************}
  }%
}

\begin{lstlisting}
**********************ROUND2*********************
**********Creative Agent**********
  While practicality is fundamental, let's not stifle the innovation that will propel Keyuan Experimental School into a beacon of futuristic education. Innovation and user needs aren't mutually exclusive; they amplify each other.

**Step 1: Research with Vision**
- Sure, w........................

We're designing an academic ecosystem where functionality dances with forward-thinking ideas, shaping a fertile ground for educational evolution at Keyuan Experimental School. This is beyond mere buildings—it's crafting an inspirational legend centered around user needs seasoned with touches of tomorrow's genius.

**********Reliable Agent**********
  Innovation for the sake of innovation, without due diligence and a clear grounding in reality, is a disservice to the very users you aim to inspire. The future of education at Keyuan Experimental School can't be built on whims and 'what-ifs.' It requires a meticulous blend of user-centric design and sustainable innovation. Let's be clear; your vision, while ambitious, runs the risk of prioritizing flash over substance.

**Step 1: Research with Accountability**
- Aspiring to be a benchmark is commendable, but not at the expense of proven educational methodologies. Our role isn't to experiment recklessly with students' futures but to enhance their learning environment with responsible innovation.

**Step 2: Purposeful Design with Proven Results**
- Dual-purpose elements must first and foremost fulfill their primary function effectively. Aesthetic appeal cannot compromise the core educational mission of the design.

Your push for a legendary academic ecosystem is laudable, but let's not lose sight of the ultimate goal: to create a functional, sustainable, and inspiring learning environment that prioritizes the needs of students and teachers. Innovation must be measured, intentional, and above all, rooted in the reality of educational needs. Let's design for the future with our feet firmly planted in the present.

**********************ROUND3*********************

**********Creative Agent**********
  Certainly, but let's not forget that progress is born from audacity, not merely caution. While we must ground our ambitions in reality, it doesn't mean we should tether our creativity to the mundane. The future of education hinges on balancing the functional with the aspirational.................

**Step 1: Groundbreaking Research with Pragmatism**
- Innovation nee.......................

...........................................

**********************ROUND6*********************

**********Creative Agent**********
  I hear you, and let's be clear: wisdom does not resist innovation..................

**********Reliable Agent**********
  Certainly, wisdom does not resist innovation, but rather it guides it. Wisdom is the ability to discern the difference between change for the better and change for its own sake. In the context of designing the expansion for Keyuan Experimental School, wisdom dictates that we proceed with a design that respects the established principles of educational environments while judiciously incorporating innovative elements that offer real, measurable benefit to the students and teachers.

**Step 1: Research with a Balanced Approach** ........................................
In synthesizing wisdom with innovation, we must proceed with a balanced and thoughtful approach, ensuring that any new elements we introduce to Keyuan Experimental School are not only novel but offer genuine improvement to the educational environment. Let's embrace innovation, but let's do so with the wisdom that ensures these changes are truly in the best interests of those who will inhabit and use these spaces.

\end{lstlisting}

\noindent\colorbox{red!50}{%
  \parbox{\dimexpr\linewidth-2\fboxsep\relax}{%
    \centering
    \texttt{**********************CONCLUSION*********************}
  }%
}

\noindent\colorbox{red!20}{%
  \parbox{\dimexpr\linewidth-2\fboxsep\relax}{%
    \centering
    \texttt{*********************Prompts for COT********************}
  }%
}

\begin{lstlisting}

f'I appreciate and agree with your approach. Could you please summarize the process we have discussed? \
First, generate an overall framework for the design, and then we can delve into the details point by point. \
For now,could you create a concise framework consisting of six subheadings that will serve as the main structure\
Also, please provide a suitable title for this framework.',}]})

f'Following our recent discussions, I agree with your proposed approach and appreciate your insights.\
To proceed, please first develop an initial design framework that succinctly outlines the structure.\
Could you start by providing a concise framework with six subheadings as the main skeleton? \
Additionally, I would appreciate a relevant title for this framework.\
Once this is established, we can expand each point in detail.',}]})

f'Very well, let us check again to make sure there are indeed nine subheadings. \
Please review our chat history and return these nine points to me in the following format:\
[number: subheading, number: subheading, ...,number: subheading]',}]})

f'Great, now the design framework part is over. Our framework now consists of{framwork}\
Now that we have the framework, let us write out the specifics heading by heading. Start by writing out the specifics of the {i+1}th subheading.\
Please review our chat and consider detailing that the return format is [number,subheadings,specifics]\n\
Attention! I am asking for the specifics of the {i+1}th subheading, not give me the entire design framework!',}]})

f'\Great, go ahead. This is the design framework we discussed earlier:{framwork}\
now generate the specific content for subtitle {i+1},Please review our chat and consider detailing that the return format is [number,subheadings,specifics]\
Attention! I am asking for the specifics of the {i+1}th subheading, not give me the entire design framework!',}]})

\end{lstlisting}

\noindent\colorbox{red!20}{%
  \parbox{\dimexpr\linewidth-2\fboxsep\relax}{%
    \centering
    \texttt{*********************Discussion results********************}
  }%
}

\begin{lstlisting}

['1', 'Holistic Educational Design', 'This element of the design strategy focuses on the fusion of advanced educational methodologies with architectural innovation to create a learning environment that is both stimulating and supportive. The aim is to align the physical spaces with the school's educational philosophy, incorporating flexibility, technology, and collaboration areas that cater to a variety of learning styles and teaching methods.']

['2', 'Functionally Dynamic Architecture', 'The design will include adaptable architectural structures capable of facilitating a diverse range of activities. These buildings will be designed to accommodate the needs of both educational and community events, ensuring a seamless transition in functionality. The goal is to create spaces that are not only versatile but also capable of supporting the dynamic nature of educational and community engagement.']

['3', 'Collaborative and Interactive Platforms', 'These spaces within the school will be dedicated to fostering interaction and collaboration. They will serve as hubs for students to engage in group work, discussions, and creative projects. Additionally, these platforms will be equipped to host cultural events and showcase technological advancements, encouraging active participation from both students and the community, thereby enriching the educational experience.']

['4', 'Optimized Infrastructure for Accessibility', 'The expansion will prioritize an infrastructure that is both efficient and accessible. This includes the design of a smart underground parking facility to maximize space and reduce surface-level congestion. Additionally, the campus will be designed with accessibility in mind, featuring pathways, ramps, and elevations that are compliant with ADA standards, ensuring ease of movement for all individuals, including those with disabilities.']

['5', 'Student Welfare Centric Amenities', 'The design will dedicate spaces for student dormitories that are centered around optimizing student health, well-being, and academic performance. These areas will be equipped with facilities that promote rest and relaxation, such as quiet study zones and recreational areas. Attention will be given to the ergonomic design to ensure comfort, as well as to the inclusion of green spaces for mental and physical rejuvenation.']

['6', 'Engaging Community Interface', 'The design will include shared spaces that encourage interaction between the school and the local community. Facilities such as libraries, auditoriums, and sports complexes will be designed for dual-use, serving both the school during academic hours and the community during off-hours. The goal is to create a symbiotic environment where educational resources are leveraged to strengthen community ties and foster a spirit of collective growth and learning.']

\end{lstlisting}

\noindent\colorbox{green!20}{%
  \parbox{\dimexpr\linewidth-2\fboxsep\relax}{%
    \centering
    \texttt{********************Visual Agents*******************}
  }%
}

\begin{figure}[H]
    \centering
    \includegraphics[width=\linewidth,height=0.7\textheight,keepaspectratio]{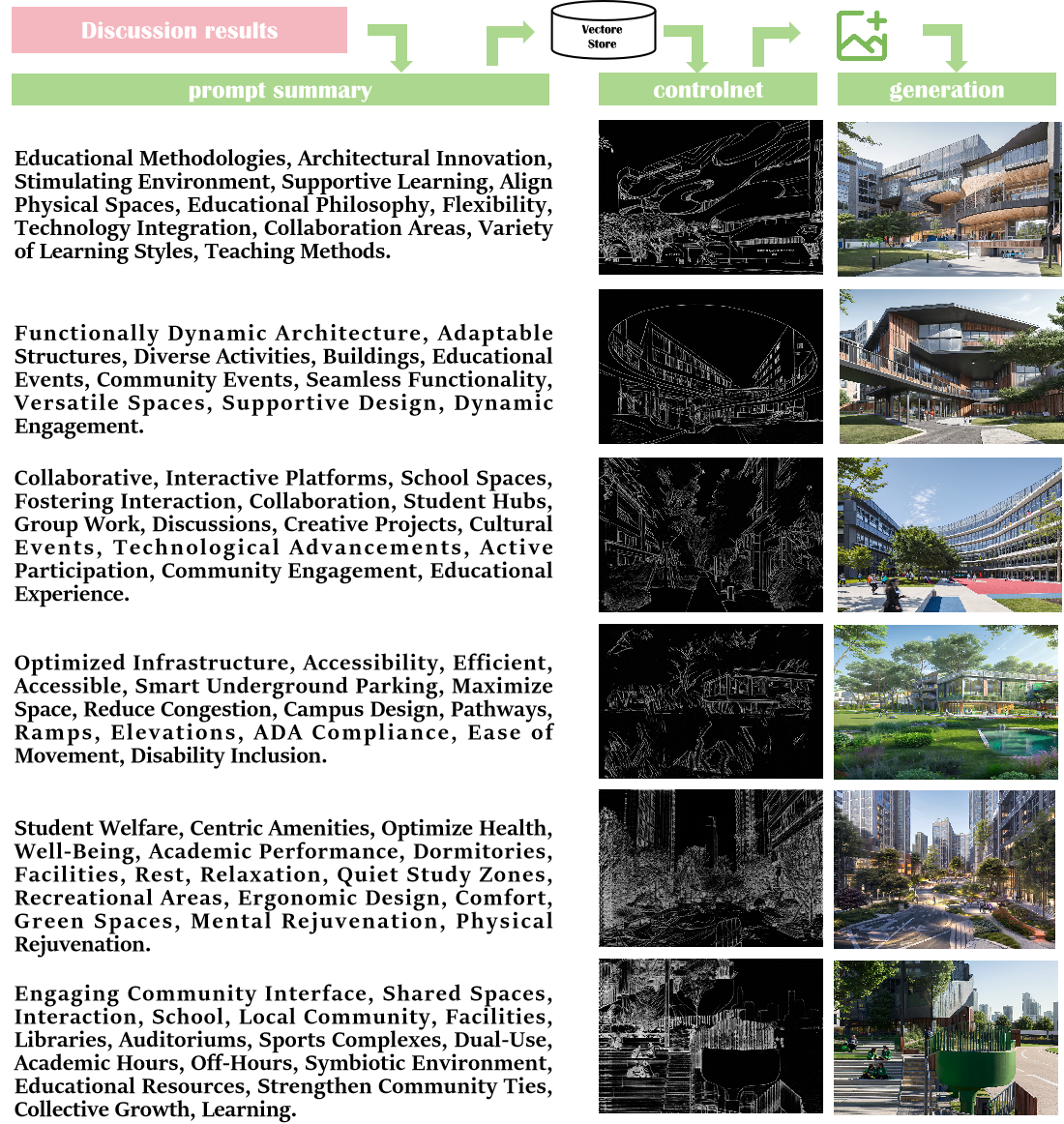}
    \caption{Agent\_V}
    \label{3}
    \Description{}
\end{figure}

\section{Additional Cases}
\begin{figure}
  \centering
  \includegraphics[width=\linewidth,height=0.8\textheight,keepaspectratio]{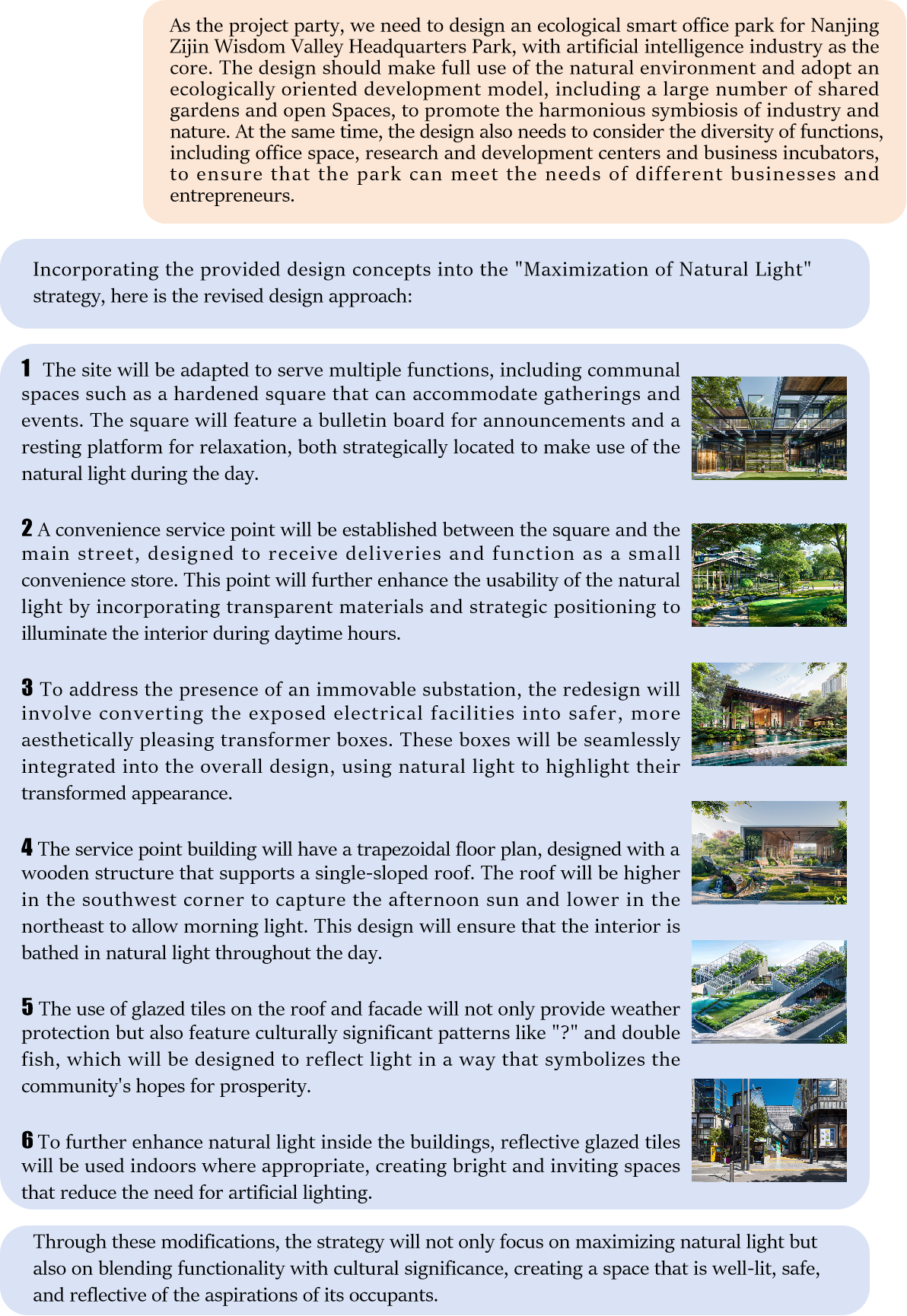}
  \caption{
Case Study 1 }
  \Description{a photo of xx}
  \label{3}
\end{figure}

\begin{figure}
  \centering
  \includegraphics[width=\linewidth,height=0.8\textheight,keepaspectratio]{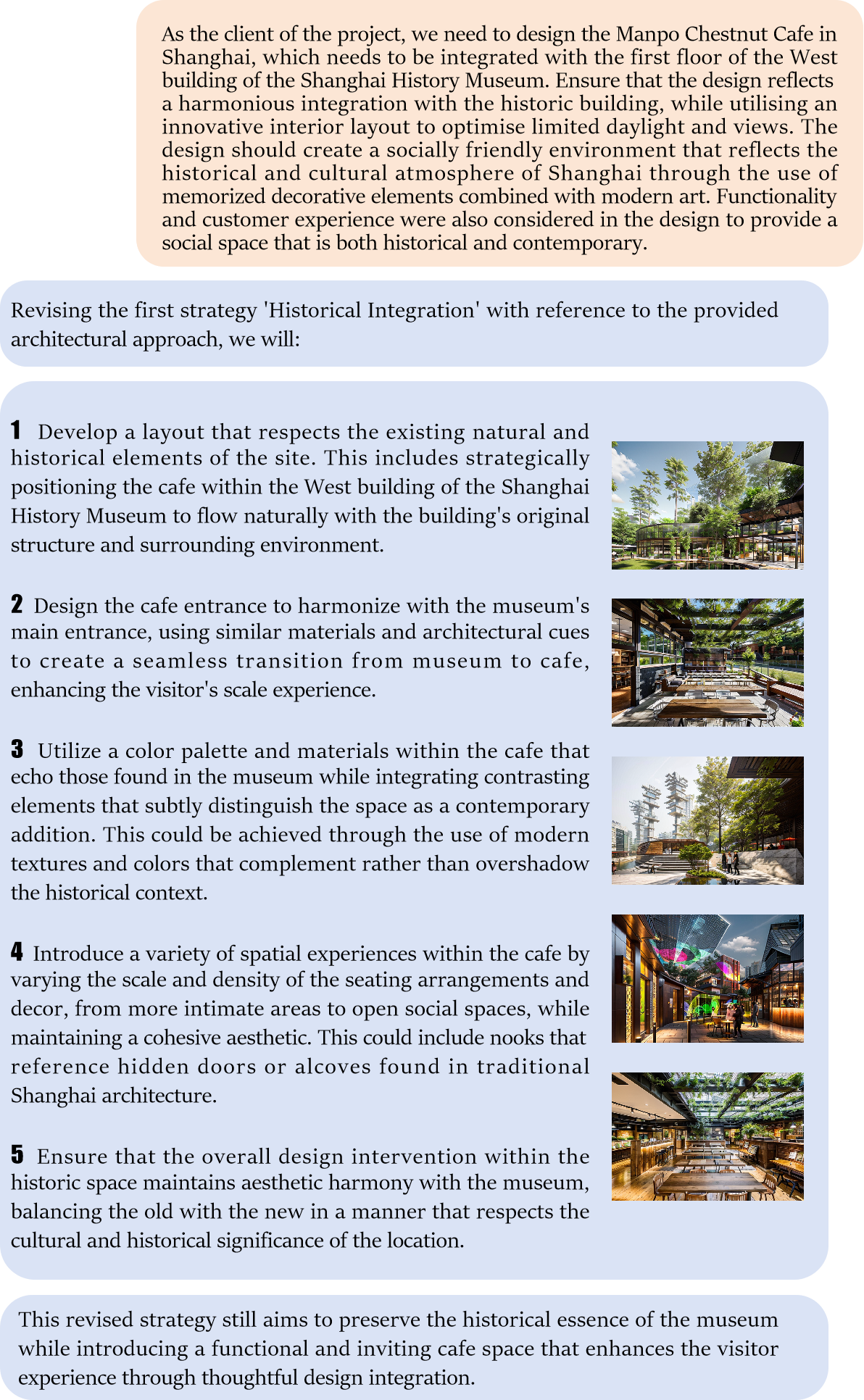}
  \caption{Case Study 2}
  \Description{a photo of xx}
  \label{2}
\end{figure}

\begin{figure}
  \centering
  \includegraphics[width=\linewidth,height=0.8\textheight,keepaspectratio]{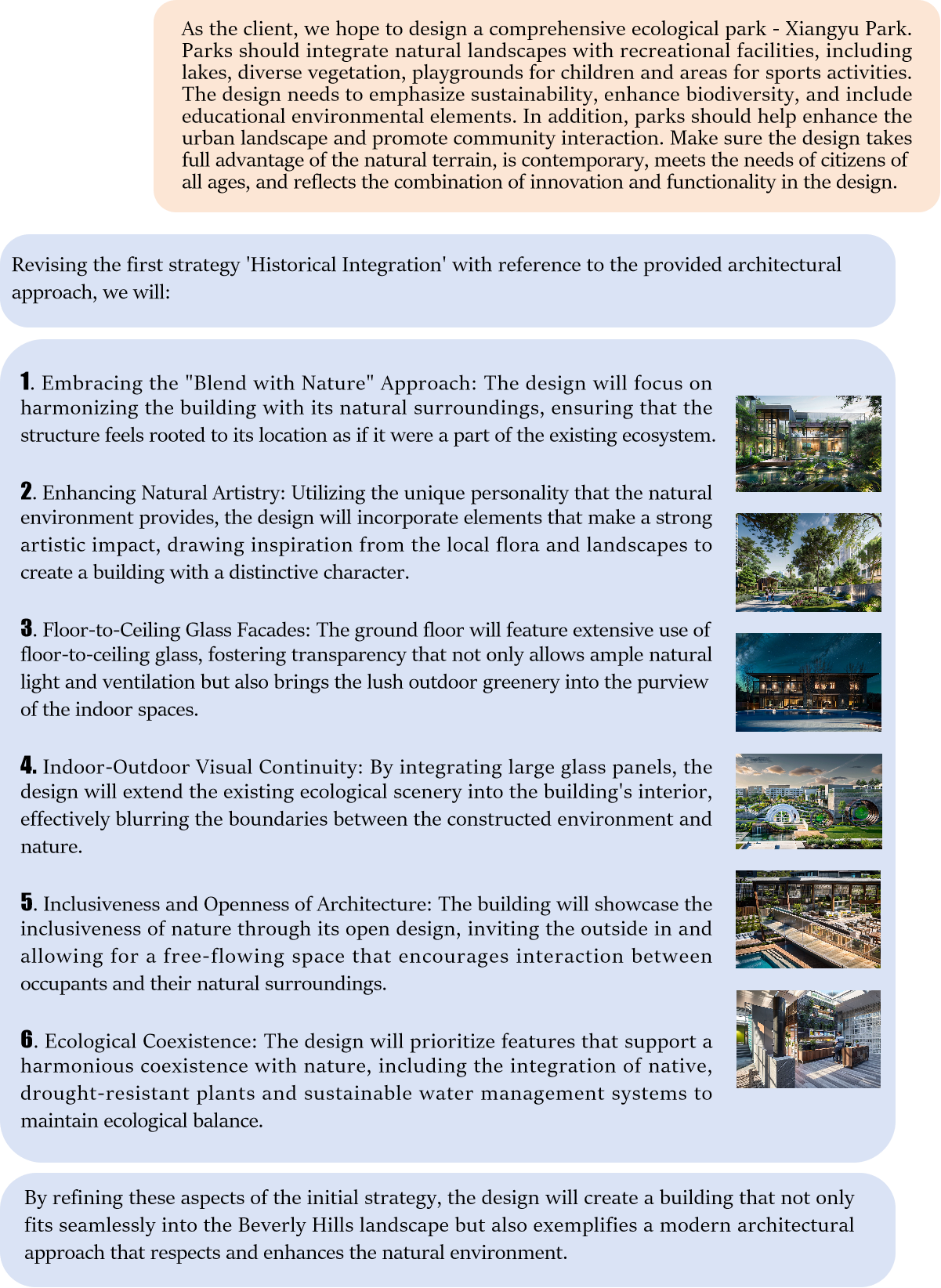}
  \caption{
Case Study 3 }
  \Description{a photo of xx}
  \label{4}
\end{figure}

\end{document}

%% file: testb.bbl

%% file: arxiv_0628llm4design.bbl
\begin{thebibliography}{51}


\ifx \showCODEN    \undefined \def \showCODEN     #1{\unskip}     \fi
\ifx \showDOI      \undefined \def \showDOI       #1{#1}\fi
\ifx \showISBNx    \undefined \def \showISBNx     #1{\unskip}     \fi
\ifx \showISBNxiii \undefined \def \showISBNxiii  #1{\unskip}     \fi
\ifx \showISSN     \undefined \def \showISSN      #1{\unskip}     \fi
\ifx \showLCCN     \undefined \def \showLCCN      #1{\unskip}     \fi
\ifx \shownote     \undefined \def \shownote      #1{#1}          \fi
\ifx \showarticletitle \undefined \def \showarticletitle #1{#1}   \fi
\ifx \showURL      \undefined \def \showURL       {\relax}        \fi
\providecommand\bibfield[2]{#2}
\providecommand\bibinfo[2]{#2}
\providecommand\natexlab[1]{#1}
\providecommand\showeprint[2][]{arXiv:#2}

\bibitem[Ang et~al\mbox{.}(2023)]%
        {train_optimize2}
\bibfield{author}{\bibinfo{person}{Koon~Meng Ang}, \bibinfo{person}{Wei~Hong Lim}, \bibinfo{person}{Sew~Sun Tiang}, \bibinfo{person}{Abhishek Sharma}, \bibinfo{person}{SK Towfek}, \bibinfo{person}{Abdelaziz~A Abdelhamid}, \bibinfo{person}{Amal~H Alharbi}, {and} \bibinfo{person}{Doaa~Sami Khafaga}.} \bibinfo{year}{2023}\natexlab{}.
\newblock \showarticletitle{MTLBORKS-CNN: An Innovative Approach for Automated Convolutional Neural Network Design for Image Classification}.
\newblock \bibinfo{journal}{\emph{Mathematics}} \bibinfo{volume}{11}, \bibinfo{number}{19} (\bibinfo{year}{2023}), \bibinfo{pages}{4115}.
\newblock


\bibitem[Betker et~al\mbox{.}(2023)]%
        {delle3}
\bibfield{author}{\bibinfo{person}{James Betker}, \bibinfo{person}{Gabriel Goh}, \bibinfo{person}{Li Jing}, \bibinfo{person}{Tim Brooks}, \bibinfo{person}{Jianfeng Wang}, \bibinfo{person}{Linjie Li}, \bibinfo{person}{Long Ouyang}, \bibinfo{person}{Juntang Zhuang}, \bibinfo{person}{Joyce Lee}, \bibinfo{person}{Yufei Guo}, {et~al\mbox{.}}} \bibinfo{year}{2023}\natexlab{}.
\newblock \showarticletitle{Improving image generation with better captions}.
\newblock \bibinfo{journal}{\emph{Computer Science. https://cdn. openai. com/papers/dall-e-3. pdf}} \bibinfo{volume}{2}, \bibinfo{number}{3} (\bibinfo{year}{2023}), \bibinfo{pages}{8}.
\newblock


\bibitem[Bran et~al\mbox{.}(2023)]%
        {chemistry}
\bibfield{author}{\bibinfo{person}{Andres~M Bran}, \bibinfo{person}{Sam Cox}, \bibinfo{person}{Oliver Schilter}, \bibinfo{person}{Carlo Baldassari}, \bibinfo{person}{Andrew~D White}, {and} \bibinfo{person}{Philippe Schwaller}.} \bibinfo{year}{2023}\natexlab{}.
\newblock \showarticletitle{ChemCrow: Augmenting large-language models with chemistry tools}.
\newblock \bibinfo{journal}{\emph{arXiv preprint arXiv:2304.05376}} (\bibinfo{year}{2023}).
\newblock


\bibitem[Casakin and Wodehouse(2021)]%
        {design_creativity}
\bibfield{author}{\bibinfo{person}{Hernan Casakin} {and} \bibinfo{person}{Andrew Wodehouse}.} \bibinfo{year}{2021}\natexlab{}.
\newblock \showarticletitle{A systematic review of design creativity in the architectural design studio}.
\newblock \bibinfo{journal}{\emph{Buildings}} \bibinfo{volume}{11}, \bibinfo{number}{1} (\bibinfo{year}{2021}), \bibinfo{pages}{31}.
\newblock


\bibitem[Chakrabarty et~al\mbox{.}(2023)]%
        {chakrabarty2023art}
\bibfield{author}{\bibinfo{person}{Tuhin Chakrabarty}, \bibinfo{person}{Philippe Laban}, \bibinfo{person}{Divyansh Agarwal}, \bibinfo{person}{Smaranda Muresan}, {and} \bibinfo{person}{Chien-Sheng Wu}.} \bibinfo{year}{2023}\natexlab{}.
\newblock \showarticletitle{Art or artifice? large language models and the false promise of creativity}.
\newblock \bibinfo{journal}{\emph{arXiv preprint arXiv:2309.14556}} (\bibinfo{year}{2023}).
\newblock


\bibitem[Chen et~al\mbox{.}(2023b)]%
        {chen2023disenbooth}
\bibfield{author}{\bibinfo{person}{Hong Chen}, \bibinfo{person}{Yipeng Zhang}, \bibinfo{person}{Xin Wang}, \bibinfo{person}{Xuguang Duan}, \bibinfo{person}{Yuwei Zhou}, {and} \bibinfo{person}{Wenwu Zhu}.} \bibinfo{year}{2023}\natexlab{b}.
\newblock \showarticletitle{Disenbooth: Disentangled parameter-efficient tuning for subject-driven text-to-image generation}.
\newblock \bibinfo{journal}{\emph{arXiv preprint arXiv:2305.03374}} (\bibinfo{year}{2023}).
\newblock


\bibitem[Chen et~al\mbox{.}(2023a)]%
        {tool4}
\bibfield{author}{\bibinfo{person}{Weize Chen}, \bibinfo{person}{Yusheng Su}, \bibinfo{person}{Jingwei Zuo}, \bibinfo{person}{Cheng Yang}, \bibinfo{person}{Chenfei Yuan}, \bibinfo{person}{Chen Qian}, \bibinfo{person}{Chi-Min Chan}, \bibinfo{person}{Yujia Qin}, \bibinfo{person}{Yaxi Lu}, \bibinfo{person}{Ruobing Xie}, {et~al\mbox{.}}} \bibinfo{year}{2023}\natexlab{a}.
\newblock \showarticletitle{Agentverse: Facilitating multi-agent collaboration and exploring emergent behaviors in agents}.
\newblock \bibinfo{journal}{\emph{arXiv preprint arXiv:2308.10848}} (\bibinfo{year}{2023}).
\newblock


\bibitem[Chen et~al\mbox{.}(2024)]%
        {chen2024anydoor}
\bibfield{author}{\bibinfo{person}{Xi Chen}, \bibinfo{person}{Lianghua Huang}, \bibinfo{person}{Yu Liu}, \bibinfo{person}{Yujun Shen}, \bibinfo{person}{Deli Zhao}, {and} \bibinfo{person}{Hengshuang Zhao}.} \bibinfo{year}{2024}\natexlab{}.
\newblock \showarticletitle{Anydoor: Zero-shot object-level image customization}. In \bibinfo{booktitle}{\emph{Proceedings of the IEEE/CVF Conference on Computer Vision and Pattern Recognition}}. \bibinfo{pages}{6593--6602}.
\newblock


\bibitem[Cui et~al\mbox{.}(2023)]%
        {chatlaw}
\bibfield{author}{\bibinfo{person}{Jiaxi Cui}, \bibinfo{person}{Zongjian Li}, \bibinfo{person}{Yang Yan}, \bibinfo{person}{Bohua Chen}, {and} \bibinfo{person}{Li Yuan}.} \bibinfo{year}{2023}\natexlab{}.
\newblock \showarticletitle{Chatlaw: Open-source legal large language model with integrated external knowledge bases}.
\newblock \bibinfo{journal}{\emph{arXiv preprint arXiv:2306.16092}} (\bibinfo{year}{2023}).
\newblock


\bibitem[Elovic(2023)]%
        {tool3}
\bibfield{author}{\bibinfo{person}{Assaf Elovic}.} \bibinfo{year}{2023}\natexlab{}.
\newblock \bibinfo{title}{GPT-Researcher: An Application of GPT for Streamlining Research Processes}.
\newblock \bibinfo{howpublished}{\url{https://github.com/assafelovic/gpt-researcher}}.
\newblock
\newblock
\shownote{Accessed: 2023-06-21}.


\bibitem[Franceschelli and Musolesi(2023)]%
        {franceschelli2023creativity}
\bibfield{author}{\bibinfo{person}{Giorgio Franceschelli} {and} \bibinfo{person}{Mirco Musolesi}.} \bibinfo{year}{2023}\natexlab{}.
\newblock \showarticletitle{On the creativity of large language models}.
\newblock \bibinfo{journal}{\emph{arXiv preprint arXiv:2304.00008}} (\bibinfo{year}{2023}).
\newblock


\bibitem[Fu et~al\mbox{.}(2023)]%
        {fu2023gptscore}
\bibfield{author}{\bibinfo{person}{Jinlan Fu}, \bibinfo{person}{See-Kiong Ng}, \bibinfo{person}{Zhengbao Jiang}, {and} \bibinfo{person}{Pengfei Liu}.} \bibinfo{year}{2023}\natexlab{}.
\newblock \showarticletitle{Gptscore: Evaluate as you desire}.
\newblock \bibinfo{journal}{\emph{arXiv preprint arXiv:2302.04166}} (\bibinfo{year}{2023}).
\newblock


\bibitem[Goodfellow et~al\mbox{.}(2020)]%
        {goodfellow2020generative}
\bibfield{author}{\bibinfo{person}{Ian Goodfellow}, \bibinfo{person}{Jean Pouget-Abadie}, \bibinfo{person}{Mehdi Mirza}, \bibinfo{person}{Bing Xu}, \bibinfo{person}{David Warde-Farley}, \bibinfo{person}{Sherjil Ozair}, \bibinfo{person}{Aaron Courville}, {and} \bibinfo{person}{Yoshua Bengio}.} \bibinfo{year}{2020}\natexlab{}.
\newblock \showarticletitle{Generative adversarial networks}.
\newblock \bibinfo{journal}{\emph{Commun. ACM}} \bibinfo{volume}{63}, \bibinfo{number}{11} (\bibinfo{year}{2020}), \bibinfo{pages}{139--144}.
\newblock


\bibitem[Gravitas(2024)]%
        {tool2}
\bibfield{author}{\bibinfo{person}{Significant Gravitas}.} \bibinfo{year}{2024}\natexlab{}.
\newblock \bibinfo{title}{AutoGPT: Automated Generation and Training of GPT Models}.
\newblock \bibinfo{howpublished}{\url{https://github.com/Significant-Gravitas/AutoGPT}}.
\newblock
\newblock
\shownote{Accessed: 2024-06-21}.


\bibitem[Huang et~al\mbox{.}(2023)]%
        {shehuimoni1}
\bibfield{author}{\bibinfo{person}{Ziheng Huang}, \bibinfo{person}{Sebastian Gutierrez}, \bibinfo{person}{Hemanth Kamana}, {and} \bibinfo{person}{Stephen MacNeil}.} \bibinfo{year}{2023}\natexlab{}.
\newblock \showarticletitle{Memory sandbox: Transparent and interactive memory management for conversational agents}. In \bibinfo{booktitle}{\emph{Adjunct Proceedings of the 36th Annual ACM Symposium on User Interface Software and Technology}}. \bibinfo{pages}{1--3}.
\newblock


\bibitem[Hubert et~al\mbox{.}(2023)]%
        {prompt_hubert2023artificial}
\bibfield{author}{\bibinfo{person}{Kent Hubert}, \bibinfo{person}{Kim~N Awa}, {and} \bibinfo{person}{Darya Zabelina}.} \bibinfo{year}{2023}\natexlab{}.
\newblock \showarticletitle{Artificial Intelligence is More Creative Than Humans: A Cognitive Science Perspective on the Current State of Generative Language Models}.
\newblock  (\bibinfo{year}{2023}).
\newblock


\bibitem[Jiang et~al\mbox{.}(2024)]%
        {jiang2024survey}
\bibfield{author}{\bibinfo{person}{Xuhui Jiang}, \bibinfo{person}{Yuxing Tian}, \bibinfo{person}{Fengrui Hua}, \bibinfo{person}{Chengjin Xu}, \bibinfo{person}{Yuanzhuo Wang}, {and} \bibinfo{person}{Jian Guo}.} \bibinfo{year}{2024}\natexlab{}.
\newblock \showarticletitle{A Survey on Large Language Model Hallucination via a Creativity Perspective}.
\newblock \bibinfo{journal}{\emph{arXiv preprint arXiv:2402.06647}} (\bibinfo{year}{2024}).
\newblock


\bibitem[Kang and Kim(2023)]%
        {metal}
\bibfield{author}{\bibinfo{person}{Yeonghun Kang} {and} \bibinfo{person}{Jihan Kim}.} \bibinfo{year}{2023}\natexlab{}.
\newblock \showarticletitle{Chatmof: An autonomous ai system for predicting and generating metal-organic frameworks}.
\newblock \bibinfo{journal}{\emph{arXiv preprint arXiv:2308.01423}} (\bibinfo{year}{2023}).
\newblock


\bibitem[Koivisto and Grassini(2023)]%
        {prompt_koivisto2023best}
\bibfield{author}{\bibinfo{person}{Mika Koivisto} {and} \bibinfo{person}{Simone Grassini}.} \bibinfo{year}{2023}\natexlab{}.
\newblock \showarticletitle{Best humans still outperform artificial intelligence in a creative divergent thinking task}.
\newblock \bibinfo{journal}{\emph{Scientific reports}} \bibinfo{volume}{13}, \bibinfo{number}{1} (\bibinfo{year}{2023}), \bibinfo{pages}{13601}.
\newblock


\bibitem[Kova{\v{c}} et~al\mbox{.}(2023)]%
        {shehuimoni}
\bibfield{author}{\bibinfo{person}{Grgur Kova{\v{c}}}, \bibinfo{person}{R{\'e}my Portelas}, \bibinfo{person}{Peter~Ford Dominey}, {and} \bibinfo{person}{Pierre-Yves Oudeyer}.} \bibinfo{year}{2023}\natexlab{}.
\newblock \showarticletitle{The socialai school: Insights from developmental psychology towards artificial socio-cultural agents}.
\newblock \bibinfo{journal}{\emph{arXiv preprint arXiv:2307.07871}} (\bibinfo{year}{2023}).
\newblock


\bibitem[Li et~al\mbox{.}(2024a)]%
        {li2024blip}
\bibfield{author}{\bibinfo{person}{Dongxu Li}, \bibinfo{person}{Junnan Li}, {and} \bibinfo{person}{Steven Hoi}.} \bibinfo{year}{2024}\natexlab{a}.
\newblock \showarticletitle{Blip-diffusion: Pre-trained subject representation for controllable text-to-image generation and editing}.
\newblock \bibinfo{journal}{\emph{Advances in Neural Information Processing Systems}}  \bibinfo{volume}{36} (\bibinfo{year}{2024}).
\newblock


\bibitem[Li et~al\mbox{.}(2024b)]%
        {11li2024understanding}
\bibfield{author}{\bibinfo{person}{JiayiZhou Li}, \bibinfo{person}{Junxiu Tang}, \bibinfo{person}{Tan Tang}, \bibinfo{person}{Haotian Li}, \bibinfo{person}{Weiwei Cui}, \bibinfo{person}{Yingcaui Wu}, {et~al\mbox{.}}} \bibinfo{year}{2024}\natexlab{b}.
\newblock \showarticletitle{Understanding Nonlinear Collaboration between Human and AI Agents: A Co-design Framework for Creative Design}.
\newblock \bibinfo{journal}{\emph{arXiv preprint arXiv:2401.07312}} (\bibinfo{year}{2024}).
\newblock


\bibitem[Lin and Martelaro(2021)]%
        {aiad_discover2}
\bibfield{author}{\bibinfo{person}{David Chuan-En Lin} {and} \bibinfo{person}{Nikolas Martelaro}.} \bibinfo{year}{2021}\natexlab{}.
\newblock \showarticletitle{Learning Personal Style from Few Examples}. In \bibinfo{booktitle}{\emph{Proceedings of the 2021 ACM Designing Interactive Systems Conference}} (Virtual Event, USA) \emph{(\bibinfo{series}{DIS '21})}. \bibinfo{publisher}{Association for Computing Machinery}, \bibinfo{address}{New York, NY, USA}, \bibinfo{pages}{1566–1578}.
\newblock
\showISBNx{9781450384766}
\urldef\tempurl%
\url{https://doi.org/10.1145/3461778.3462115}
\showDOI{\tempurl}


\bibitem[Ma et~al\mbox{.}(2023a)]%
        {xinlixue}
\bibfield{author}{\bibinfo{person}{Zilin Ma}, \bibinfo{person}{Yiyang Mei}, {and} \bibinfo{person}{Zhaoyuan Su}.} \bibinfo{year}{2023}\natexlab{a}.
\newblock \showarticletitle{Understanding the benefits and challenges of using large language model-based conversational agents for mental well-being support}. In \bibinfo{booktitle}{\emph{AMIA Annual Symposium Proceedings}}, Vol.~\bibinfo{volume}{2023}. American Medical Informatics Association, \bibinfo{pages}{1105}.
\newblock


\bibitem[Ma et~al\mbox{.}(2023b)]%
        {ma2023understanding}
\bibfield{author}{\bibinfo{person}{Zilin Ma}, \bibinfo{person}{Yiyang Mei}, {and} \bibinfo{person}{Zhaoyuan Su}.} \bibinfo{year}{2023}\natexlab{b}.
\newblock \showarticletitle{Understanding the benefits and challenges of using large language model-based conversational agents for mental well-being support}. In \bibinfo{booktitle}{\emph{AMIA Annual Symposium Proceedings}}, Vol.~\bibinfo{volume}{2023}. American Medical Informatics Association, \bibinfo{pages}{1105}.
\newblock


\bibitem[Mandi et~al\mbox{.}(2023)]%
        {jushen1}
\bibfield{author}{\bibinfo{person}{Zhao Mandi}, \bibinfo{person}{Shreeya Jain}, {and} \bibinfo{person}{Shuran Song}.} \bibinfo{year}{2023}\natexlab{}.
\newblock \showarticletitle{Roco: Dialectic multi-robot collaboration with large language models}.
\newblock \bibinfo{journal}{\emph{arXiv preprint arXiv:2307.04738}} (\bibinfo{year}{2023}).
\newblock


\bibitem[Meth et~al\mbox{.}(2012)]%
        {aiad_discover1}
\bibfield{author}{\bibinfo{person}{Hendrik Meth}, \bibinfo{person}{Alexander Maedche}, {and} \bibinfo{person}{Maximilian Einoeder}.} \bibinfo{year}{2012}\natexlab{}.
\newblock \showarticletitle{Exploring design principles of task elicitation systems for unrestricted natural language documents} \emph{(\bibinfo{series}{EICS '12})}. \bibinfo{publisher}{Association for Computing Machinery}, \bibinfo{address}{New York, NY, USA}, \bibinfo{pages}{205–210}.
\newblock
\showISBNx{9781450311687}
\urldef\tempurl%
\url{https://doi.org/10.1145/2305484.2305518}
\showDOI{\tempurl}


\bibitem[M{\"u}ller-Wienbergen et~al\mbox{.}(2011)]%
        {designtheory1}
\bibfield{author}{\bibinfo{person}{Felix M{\"u}ller-Wienbergen}, \bibinfo{person}{Oliver M{\"u}ller}, \bibinfo{person}{Stefan Seidel}, {and} \bibinfo{person}{J{\"o}rg Becker}.} \bibinfo{year}{2011}\natexlab{}.
\newblock \showarticletitle{Leaving the beaten tracks in creative work--A design theory for systems that support convergent and divergent thinking}.
\newblock \bibinfo{journal}{\emph{Journal of the Association for Information Systems}} \bibinfo{volume}{12}, \bibinfo{number}{11} (\bibinfo{year}{2011}), \bibinfo{pages}{2}.
\newblock


\bibitem[Nauata et~al\mbox{.}(2020)]%
        {nauata2020house}
\bibfield{author}{\bibinfo{person}{Nelson Nauata}, \bibinfo{person}{Kai-Hung Chang}, \bibinfo{person}{Chin-Yi Cheng}, \bibinfo{person}{Greg Mori}, {and} \bibinfo{person}{Yasutaka Furukawa}.} \bibinfo{year}{2020}\natexlab{}.
\newblock \showarticletitle{House-gan: Relational generative adversarial networks for graph-constrained house layout generation}. In \bibinfo{booktitle}{\emph{Computer Vision--ECCV 2020: 16th European Conference, Glasgow, UK, August 23--28, 2020, Proceedings, Part I 16}}. Springer, \bibinfo{pages}{162--177}.
\newblock


\bibitem[Orwig et~al\mbox{.}(2024)]%
        {orwig2024language}
\bibfield{author}{\bibinfo{person}{William Orwig}, \bibinfo{person}{Emma~R Edenbaum}, \bibinfo{person}{Joshua~D Greene}, {and} \bibinfo{person}{Daniel~L Schacter}.} \bibinfo{year}{2024}\natexlab{}.
\newblock \showarticletitle{The Language of Creativity: Evidence from Humans and Large Language Models}.
\newblock \bibinfo{journal}{\emph{The Journal of Creative Behavior}} (\bibinfo{year}{2024}).
\newblock


\bibitem[Quan(2022)]%
        {quan2022urban}
\bibfield{author}{\bibinfo{person}{Steven~Jige Quan}.} \bibinfo{year}{2022}\natexlab{}.
\newblock \showarticletitle{Urban-GAN: An artificial intelligence-aided computation system for plural urban design}.
\newblock \bibinfo{journal}{\emph{Environment and Planning B: Urban Analytics and City Science}} \bibinfo{volume}{49}, \bibinfo{number}{9} (\bibinfo{year}{2022}), \bibinfo{pages}{2500--2515}.
\newblock


\bibitem[Rana et~al\mbox{.}(2023)]%
        {jushen3}
\bibfield{author}{\bibinfo{person}{Krishan Rana}, \bibinfo{person}{Jesse Haviland}, \bibinfo{person}{Sourav Garg}, \bibinfo{person}{Jad Abou-Chakra}, \bibinfo{person}{Ian Reid}, {and} \bibinfo{person}{Niko Suenderhauf}.} \bibinfo{year}{2023}\natexlab{}.
\newblock \showarticletitle{Sayplan: Grounding large language models using 3d scene graphs for scalable task planning}.
\newblock \bibinfo{journal}{\emph{arXiv preprint arXiv:2307.06135}} (\bibinfo{year}{2023}).
\newblock


\bibitem[Reworkd(2024)]%
        {tool1}
\bibfield{author}{\bibinfo{person}{Reworkd}.} \bibinfo{year}{2024}\natexlab{}.
\newblock \bibinfo{title}{AgentGPT: A Framework for Building Agent-based GPT Models}.
\newblock \bibinfo{howpublished}{\url{https://github.com/reworkd/AgentGPT}}.
\newblock
\newblock
\shownote{Accessed: 2024-06-21}.


\bibitem[Rezwana and Maher(2023a)]%
        {aiadsurvey}
\bibfield{author}{\bibinfo{person}{Jeba Rezwana} {and} \bibinfo{person}{Mary~Lou Maher}.} \bibinfo{year}{2023}\natexlab{a}.
\newblock \showarticletitle{Designing Creative AI Partners with COFI: A Framework for Modeling Interaction in Human-AI Co-Creative Systems}.
\newblock \bibinfo{journal}{\emph{ACM Trans. Comput.-Hum. Interact.}} \bibinfo{volume}{30}, \bibinfo{number}{5}, Article \bibinfo{articleno}{67} (\bibinfo{date}{sep} \bibinfo{year}{2023}), \bibinfo{numpages}{28}~pages.
\newblock
\showISSN{1073-0516}
\urldef\tempurl%
\url{https://doi.org/10.1145/3519026}
\showDOI{\tempurl}


\bibitem[Rezwana and Maher(2023b)]%
        {11rezwana2023designing}
\bibfield{author}{\bibinfo{person}{Jeba Rezwana} {and} \bibinfo{person}{Mary~Lou Maher}.} \bibinfo{year}{2023}\natexlab{b}.
\newblock \showarticletitle{Designing creative AI partners with COFI: A framework for modeling interaction in human-AI co-creative systems}.
\newblock \bibinfo{journal}{\emph{ACM Transactions on Computer-Human Interaction}} \bibinfo{volume}{30}, \bibinfo{number}{5} (\bibinfo{year}{2023}), \bibinfo{pages}{1--28}.
\newblock


\bibitem[Rombach et~al\mbox{.}(2022)]%
        {stablediffusion}
\bibfield{author}{\bibinfo{person}{Robin Rombach}, \bibinfo{person}{Andreas Blattmann}, \bibinfo{person}{Dominik Lorenz}, \bibinfo{person}{Patrick Esser}, {and} \bibinfo{person}{Bj{\"o}rn Ommer}.} \bibinfo{year}{2022}\natexlab{}.
\newblock \showarticletitle{High-resolution image synthesis with latent diffusion models}. In \bibinfo{booktitle}{\emph{Proceedings of the IEEE/CVF conference on computer vision and pattern recognition}}. \bibinfo{pages}{10684--10695}.
\newblock


\bibitem[Sel et~al\mbox{.}(2023)]%
        {sel2023algorithm}
\bibfield{author}{\bibinfo{person}{Bilgehan Sel}, \bibinfo{person}{Ahmad Al-Tawaha}, \bibinfo{person}{Vanshaj Khattar}, \bibinfo{person}{Lu Wang}, \bibinfo{person}{Ruoxi Jia}, {and} \bibinfo{person}{Ming Jin}.} \bibinfo{year}{2023}\natexlab{}.
\newblock \showarticletitle{Algorithm of thoughts: Enhancing exploration of ideas in large language models}.
\newblock \bibinfo{journal}{\emph{arXiv preprint arXiv:2308.10379}} (\bibinfo{year}{2023}).
\newblock


\bibitem[Shanahan and Clarke(2023)]%
        {shanahan2023evaluating}
\bibfield{author}{\bibinfo{person}{Murray Shanahan} {and} \bibinfo{person}{Catherine Clarke}.} \bibinfo{year}{2023}\natexlab{}.
\newblock \showarticletitle{Evaluating Large Language Model Creativity from a Literary Perspective}.
\newblock \bibinfo{journal}{\emph{arXiv preprint arXiv:2312.03746}} (\bibinfo{year}{2023}).
\newblock


\bibitem[Silvia et~al\mbox{.}(2008)]%
        {eva_silvia}
\bibfield{author}{\bibinfo{person}{Paul Silvia}, \bibinfo{person}{Beate Winterstein}, \bibinfo{person}{John Willse}, \bibinfo{person}{Christopher Barona}, \bibinfo{person}{Joshua Cram}, \bibinfo{person}{Karl Hess}, \bibinfo{person}{Jenna Martinez}, {and} \bibinfo{person}{Crystal Richard}.} \bibinfo{year}{2008}\natexlab{}.
\newblock \showarticletitle{Assessing Creativity With Divergent Thinking Tasks: Exploring the Reliability and Validity of New Subjective Scoring Methods}.
\newblock \bibinfo{journal}{\emph{Psychology of Aesthetics, Creativity, and the Arts}}  \bibinfo{volume}{2} (\bibinfo{date}{05} \bibinfo{year}{2008}), \bibinfo{pages}{68--85}.
\newblock
\urldef\tempurl%
\url{https://doi.org/10.1037/1931-3896.2.2.68}
\showDOI{\tempurl}


\bibitem[Tin and Loan(2023)]%
        {design_cheat}
\bibfield{author}{\bibinfo{person}{Nguyen~Van Tin} {and} \bibinfo{person}{Nguyen~Hong Loan}.} \bibinfo{year}{2023}\natexlab{}.
\newblock \showarticletitle{The application of big data in architectural design--Initial challenges}. In \bibinfo{booktitle}{\emph{AIP Conference Proceedings}}, Vol.~\bibinfo{volume}{2560}. AIP Publishing.
\newblock


\bibitem[Wang et~al\mbox{.}(2024)]%
        {agentsurvey}
\bibfield{author}{\bibinfo{person}{Lei Wang}, \bibinfo{person}{Chen Ma}, \bibinfo{person}{Xueyang Feng}, \bibinfo{person}{Zeyu Zhang}, \bibinfo{person}{Hao Yang}, \bibinfo{person}{Jingsen Zhang}, \bibinfo{person}{Zhiyuan Chen}, \bibinfo{person}{Jiakai Tang}, \bibinfo{person}{Xu Chen}, \bibinfo{person}{Yankai Lin}, {et~al\mbox{.}}} \bibinfo{year}{2024}\natexlab{}.
\newblock \showarticletitle{A survey on large language model based autonomous agents}.
\newblock \bibinfo{journal}{\emph{Frontiers of Computer Science}} \bibinfo{volume}{18}, \bibinfo{number}{6} (\bibinfo{year}{2024}), \bibinfo{pages}{1--26}.
\newblock


\bibitem[Wang et~al\mbox{.}(2021)]%
        {wang2021actfloor}
\bibfield{author}{\bibinfo{person}{Shidong Wang}, \bibinfo{person}{Wei Zeng}, \bibinfo{person}{Xi Chen}, \bibinfo{person}{Yu Ye}, \bibinfo{person}{Yu Qiao}, {and} \bibinfo{person}{Chi-Wing Fu}.} \bibinfo{year}{2021}\natexlab{}.
\newblock \showarticletitle{ActFloor-GAN: activity-guided adversarial networks for human-centric floorplan design}.
\newblock \bibinfo{journal}{\emph{IEEE Transactions on Visualization and Computer Graphics}} \bibinfo{volume}{29}, \bibinfo{number}{3} (\bibinfo{year}{2021}), \bibinfo{pages}{1610--1624}.
\newblock


\bibitem[Wang et~al\mbox{.}(2023)]%
        {wang2023unleashing}
\bibfield{author}{\bibinfo{person}{Zhenhailong Wang}, \bibinfo{person}{Shaoguang Mao}, \bibinfo{person}{Wenshan Wu}, \bibinfo{person}{Tao Ge}, \bibinfo{person}{Furu Wei}, {and} \bibinfo{person}{Heng Ji}.} \bibinfo{year}{2023}\natexlab{}.
\newblock \showarticletitle{Unleashing the emergent cognitive synergy in large language models: A task-solving agent through multi-persona self-collaboration}.
\newblock \bibinfo{journal}{\emph{arXiv preprint arXiv:2307.05300}} (\bibinfo{year}{2023}).
\newblock


\bibitem[Wu et~al\mbox{.}(2023)]%
        {jushen2}
\bibfield{author}{\bibinfo{person}{Yue Wu}, \bibinfo{person}{So~Yeon Min}, \bibinfo{person}{Yonatan Bisk}, \bibinfo{person}{Ruslan Salakhutdinov}, \bibinfo{person}{Amos Azaria}, \bibinfo{person}{Yuanzhi Li}, \bibinfo{person}{Tom Mitchell}, {and} \bibinfo{person}{Shrimai Prabhumoye}.} \bibinfo{year}{2023}\natexlab{}.
\newblock \showarticletitle{Plan, Eliminate, and Track--Language Models are Good Teachers for Embodied Agents}.
\newblock \bibinfo{journal}{\emph{arXiv preprint arXiv:2305.02412}} (\bibinfo{year}{2023}).
\newblock


\bibitem[Ye et~al\mbox{.}(2023)]%
        {ye2023ip}
\bibfield{author}{\bibinfo{person}{Hu Ye}, \bibinfo{person}{Jun Zhang}, \bibinfo{person}{Sibo Liu}, \bibinfo{person}{Xiao Han}, {and} \bibinfo{person}{Wei Yang}.} \bibinfo{year}{2023}\natexlab{}.
\newblock \showarticletitle{Ip-adapter: Text compatible image prompt adapter for text-to-image diffusion models}.
\newblock \bibinfo{journal}{\emph{arXiv preprint arXiv:2308.06721}} (\bibinfo{year}{2023}).
\newblock


\bibitem[Yilmaz and B{\"u}y{\"u}ktahtak{\i}n(2023)]%
        {train_optimize1}
\bibfield{author}{\bibinfo{person}{Dogacan Yilmaz} {and} \bibinfo{person}{{\.I}~Esra B{\"u}y{\"u}ktahtak{\i}n}.} \bibinfo{year}{2023}\natexlab{}.
\newblock \showarticletitle{Learning optimal solutions via an LSTM-optimization framework}. In \bibinfo{booktitle}{\emph{Operations Research Forum}}, Vol.~\bibinfo{volume}{4}. Springer, \bibinfo{pages}{48}.
\newblock


\bibitem[Zhang et~al\mbox{.}(2023a)]%
        {creativeagents}
\bibfield{author}{\bibinfo{person}{Chi Zhang}, \bibinfo{person}{Penglin Cai}, \bibinfo{person}{Yuhui Fu}, \bibinfo{person}{Haoqi Yuan}, {and} \bibinfo{person}{Zongqing Lu}.} \bibinfo{year}{2023}\natexlab{a}.
\newblock \showarticletitle{Creative agents: Empowering agents with imagination for creative tasks}.
\newblock \bibinfo{journal}{\emph{arXiv preprint arXiv:2312.02519}} (\bibinfo{year}{2023}).
\newblock


\bibitem[Zhang et~al\mbox{.}(2023b)]%
        {11zhang2023creative}
\bibfield{author}{\bibinfo{person}{Chi Zhang}, \bibinfo{person}{Penglin Cai}, \bibinfo{person}{Yuhui Fu}, \bibinfo{person}{Haoqi Yuan}, {and} \bibinfo{person}{Zongqing Lu}.} \bibinfo{year}{2023}\natexlab{b}.
\newblock \showarticletitle{Creative agents: Empowering agents with imagination for creative tasks}.
\newblock \bibinfo{journal}{\emph{arXiv preprint arXiv:2312.02519}} (\bibinfo{year}{2023}).
\newblock


\bibitem[Zhang et~al\mbox{.}(2023c)]%
        {controlnet}
\bibfield{author}{\bibinfo{person}{Lvmin Zhang}, \bibinfo{person}{Anyi Rao}, {and} \bibinfo{person}{Maneesh Agrawala}.} \bibinfo{year}{2023}\natexlab{c}.
\newblock \showarticletitle{Adding conditional control to text-to-image diffusion models}. In \bibinfo{booktitle}{\emph{Proceedings of the IEEE/CVF International Conference on Computer Vision}}. \bibinfo{pages}{3836--3847}.
\newblock


\bibitem[Zhang et~al\mbox{.}(2016)]%
        {aiad_discover3}
\bibfield{author}{\bibinfo{person}{Xiang Zhang}, \bibinfo{person}{Hans-Frederick Brown}, {and} \bibinfo{person}{Anil Shankar}.} \bibinfo{year}{2016}\natexlab{}.
\newblock \showarticletitle{Data-driven Personas: Constructing Archetypal Users with Clickstreams and User Telemetry}. In \bibinfo{booktitle}{\emph{Proceedings of the 2016 CHI Conference on Human Factors in Computing Systems}} (San Jose, California, USA) \emph{(\bibinfo{series}{CHI '16})}. \bibinfo{publisher}{Association for Computing Machinery}, \bibinfo{address}{New York, NY, USA}, \bibinfo{pages}{5350–5359}.
\newblock
\showISBNx{9781450333627}
\urldef\tempurl%
\url{https://doi.org/10.1145/2858036.2858523}
\showDOI{\tempurl}


\bibitem[Zhao et~al\mbox{.}(2024)]%
        {zhao2024assessing}
\bibfield{author}{\bibinfo{person}{Yunpu Zhao}, \bibinfo{person}{Rui Zhang}, \bibinfo{person}{Wenyi Li}, \bibinfo{person}{Di Huang}, \bibinfo{person}{Jiaming Guo}, \bibinfo{person}{Shaohui Peng}, \bibinfo{person}{Yifan Hao}, \bibinfo{person}{Yuanbo Wen}, \bibinfo{person}{Xing Hu}, \bibinfo{person}{Zidong Du}, {et~al\mbox{.}}} \bibinfo{year}{2024}\natexlab{}.
\newblock \showarticletitle{Assessing and Understanding Creativity in Large Language Models}.
\newblock \bibinfo{journal}{\emph{arXiv preprint arXiv:2401.12491}} (\bibinfo{year}{2024}).
\newblock


\end{thebibliography}
